\documentclass[aps,pra,twocolumn,superscriptaddress,showpacs,showkeys]{revtex4-1}

\usepackage{amsmath}
\usepackage{amssymb}
\usepackage{braket}
\usepackage{color}
\usepackage{dsfont}
\usepackage{graphics}
\usepackage{graphicx}
\usepackage{slashed}
\usepackage[hidelinks]{hyperref}
\usepackage{todonotes}
\usepackage{eufrak}

\renewcommand{\vec}[1]{\boldsymbol{#1}}
\newcommand{\T}{\textrm{T}}

\begin{document}

\title{Spin-dependent two-photon Bragg scattering in the Kapitza-Dirac effect}

\author{Sven Ahrens}
\email{ahrens@shnu.edu.cn}
\affiliation{Department of Physics, Shanghai Normal University, Shanghai 200234, China}
\author{Zhenfeng Liang}
\affiliation{Department of Physics, Shanghai Normal University, Shanghai 200234, China}
\author{Tilen \v Cade\v z}
\affiliation{Center for Theoretical Physics of Complex Systems, Institute for Basic Science (IBS), Daejeon 34126, Republic of Korea}
\affiliation{Center of Physics of University of Minho and University of Porto, P-4169-007 Oporto, Portugal}
\affiliation{CAS Key Laboratory of Theoretical Physics, Institute of Theoretical Physics, Chinese Academy of Sciences, Beijing 100193, China}
\author{Baifei Shen}
\email{bfshen@shnu.edu.cn}
\affiliation{Department of Physics, Shanghai Normal University, Shanghai 200234, China}

\date{\today}

\begin{abstract}
We present the possibility of spin-dependent Kapitza-Dirac scattering based on a two-photon interaction only. The interaction scheme is inspired from a Compton scattering process, for which we explicitly show the mathematical correspondence to the spin-dynamics of an electron diffraction process in a standing light wave. The spin effect has the advantage that it already appears in a Bragg scattering setup with arbitrary low field amplitudes, for which we have estimated the diffraction count rate in a realistic experimental setup at available X-ray free-electron laser facilities.
\end{abstract}


\maketitle

\section{Introduction\label{sec:introduction}}

The spin is an intrinsic angular momentum of every elementary particle
\footnote{More specifically, the spin is an intrinsic angular momentum of every elementary particle, where different elementary particles have different values of total spin angular momentum. In the exceptional case of the Higgs boson, the total spin angular momentum is 0, being therefore a spinless scalar particle. The other known elementary particles have a total spin angular momentum larger than 0.}. While from the theoretical point of view one would identify the spin as a byproduct of the quantization procedure of relativistic wave functions, one might in a classical picture imagine the spin as a tiny spinning sphere. This view might be intuitive, but should be considered as technically incorrect. Nevertheless, one is associating a magnetic moment with a magnetic dipole to the electron, which in the classical imagination of a charged, spinning sphere would be the `handle' to interact with the electron spin. More formally, intrinsic angular momentum is characterized by ``unitary representations of the inhomogeneous Lorentz group'', according to Wigner \cite{Wigner_1939_representation_theory,Weinberg_1995_quantum_theory_of_fields}. From the historical perspective, it seems that the electron spin was initially rather an implication from the need for a consistent explanation for the atomic structure, as well as from spectroscopic observations \cite{Morrison_2007_spin}, with  a first explicit experimental indication from the Stern-Gerlach experiment \cite{Stern_Gerlach_1922_1,Stern_Gerlach_1922_2,Stern_Gerlach_1922_3}.

Within the scientific applications of present times, spin-dependent electron interaction appears commonly in photo-emission \cite{Kirschner_1981_spin_resolved_photo_emission,Maruyama_1991_spin_polarized_photo_emission}, such that interesting applications like  spin- and angle-resolved photoemission spectroscopy (SARPES) \cite{Meier_2008_sarpes,He_2010_sarpes,Bentmann_2011_sarpes} is possible, with even recording spin-resolved band structure \cite{Kutnaykhov_2015_band_structure,Elmers_2016_band_structure,Noguchi_2017_spin_band_structure,tang_2012_spin_talbot_effect}. However, these examples have in common that they are bound state systems, in which the electron is not free from interactions with its environment. For isolated electrons, which propagate freely in space, Wolfgang Pauli has already argued in 1932 that an electron interaction with electro-magnetic fields cannot be sensitive to the electron spin in terms of a concept of classical trajectories \cite{pauli_1932_classical_trajectories}. The reason is that the Stern-Gerlach experiment has been carried out with electrically neutral silver atoms instead of charged electrons. For charged particles, however, the Lorentz force from the magnetic field in the Stern-Gerlach experiment requires a precise knowledge of the electron's initial position and momentum, which is in conflict with the Heisenberg uncertainty relation. Therefore, a common assumption is that ``it is impossible, by means of a Stern-Gerlach experiment, to determine the magnetic moment of a free electron'' \cite{mott_massey_1965_atomic_collisions} and that ``Conventional spin filters, the prototype of which is the Stern-Gerlach magnet, do not work with free electrons.'' \cite{kessler_1976_polarized_electrons}. Nevertheless, proposals for a longitudinal setup of the Stern-Gerlach experiment with electrons exist \cite{Batelaan_1997_electron_stern-gerlach_first,rutherford_grobe_1998_electron_stern-gerlach_comment}, for a ``minimum-spreading longitudinal configuration'' \cite{Gallup_2001_electron_stern-gerlach_quantum}. Also, random spin-flips can be induced by radio frequency field injection and thermal radiation at an electron in a Penning trap \cite{Dehmelt_1988_stern_gerlach_zeitschrift_physik,Dehmelt_1990_stern_gerlach_science}.

Electron diffraction in standing light waves, as first proposed by Kapitza and Dirac \cite{kapitza_dirac_1933_proposal} (see also \cite{Federov_1980_multiphoton_stimulated_compton_scattering,gush_gush_1971_higher_order_kapitza_dirac_scattering,Efremov_1999_classical_and_quantum_KDE,Efremov_2000_wavepacket_theory_KDE,Smirnova_2004_diffraction_without_grating,Zhang_2004_theory_KDE}) could be a way to establish a controlled and explicit spin-dependent interaction of electrons with electro-magnetic fields only. A spin-\emph{independent} Kapitza-Dirac effect has already been experimentally demonstrated for atoms \cite{gould_1986_atoms_diffraction_regime,martin_1988_atoms_bragg_regime} and also electrons in a strong \cite{Bucksbaum_1988_electron_diffraction_regime} and weak \cite{Freimund_Batelaan_2001_KDE_first,Freimund_Batelaan_2002_KDE_detection_PRL} interaction regime. Concerning `strong' and `weak' interaction regimes, we follow a characterization from Batelaan \cite{batelaan_2000_KDE_first,batelaan_2007_RMP_KDE}, where the recoil shift $\epsilon$ (corresponding to the spacing of the kinetic energy of the different electron diffraction orders) is compared to the ponderomotive amplitude $V_0$ of the standing light wave. The system is in the Bragg regime (weak interaction), if $\epsilon\gg V_0$ and in the diffraction regime for $\epsilon\ll V_0$ (strong interaction). Spin effects \cite{Batelaan_2003_MSGE,rosenstein_2004_first_KDE_spin_calculation,ahrens_bauke_2012_spin-kde,ahrens_bauke_2013_relativistic_KDE,bauke_ahrens_2014_spin_precession_1,bauke_ahrens_2014_spin_precession_2,erhard_bauke_2015_spin,McGregor_Batelaan_2015_two_color_spin,dellweg_awwad_mueller_2016_spin-dynamics_bichromatic_laser_fields} and also spin-dependent diffraction \cite{dellweg_mueller_2016_interferometric_spin-polarizer,dellweg_mueller_extended_KDE_calculations,ahrens_2017_spin_filter,ebadati_2018_four_photon_KDE,ebadati_2019_n_photon_KDE} (ie. sorting of electrons according to their spin state) has been discussed theoretically for the Kapitza-Dirac effect. While the original proposal from Kapitza and Dirac considers a two-photon momentum transfer, higher order photon scattering is possible \cite{ahrens_bauke_2012_spin-kde,ahrens_bauke_2013_relativistic_KDE,McGregor_Batelaan_2015_two_color_spin,dellweg_awwad_mueller_2016_spin-dynamics_bichromatic_laser_fields,dellweg_mueller_2016_interferometric_spin-polarizer,dellweg_mueller_extended_KDE_calculations,ebadati_2018_four_photon_KDE,Kozak_2018_ponderomotive_KDE,ebadati_2019_n_photon_KDE} but might be suppressed for the case of a weak ponderomotive amplitude of the standing wave light field in the Bragg regime. Therefore, possible implementation difficulties of spin-dependent electron diffraction could arise for the case of a higher number of interacting photons or the necessity to wait for larger fractions of a Rabi cycle of the electron quantum state transition, which could hinder the observation of such higher order photon interactions. A further discussion about spin-dependent electron diffraction scenarios in laser fields is carried out in the outlook section \ref{sec:conclusion_and_outlook} at the end of this article. We also point out other theoretical investigations of electron spin dynamics in strong laser fields \cite{panek_2002_laser-induced_compton_scattering,ivanov_2004_polarization_effects_in_strong_laser,boca_florescu_2009_nonlinear_compton_scattering,krajewska_kaminski_2013_spin_effects_in_compton_scattering,Skoromnik_2013_spin_revivial,King_2015_double_compton_scattering_constand_crossed_field,Sorbo_2017_intense_spin_polarization,Sorbo_2018_theory_electron_polarization,li_2019_ultra_relativistic_polarization,Chen_2019_polarized_positron_beams,Wen_Tamburini_2019_polarized_kiloampere_electron_beams,Fu_2019_three_dimensional_spin_dirac_solutions} as well as spin-independent electron diffraction scenarios with a controlled phase-space construction \cite{Kling_2015_qfel_quantum_regime,Kling_2016_relativistic_qfel,Debus_2019_qfel_realization,Kling_2019_qfel_gain,Carmesin_2020_phase_space_dynamics_free_electron_laser}.

In this article we discuss spin-dependent Kapitza-Dirac diffraction, featuring a two photon interaction (first feature), which takes place in a Bragg scattering scenario (second feature). In this context, the term ``two photon interaction'' means that the electron absorbs and emits one photon in a classical view of the interaction. The second feature ``Bragg scattering scenario'' implies that coherent population transfer between the incoming and diffracted mode allows for the statistical observation of the effect at theoretically arbitrary low field amplitudes. The approach is inspired by a previous work of one of us, which is investigating spin properties in Compton scattering \cite{ahrens_2017_spin_non_conservation}. Accordingly, the effect can be  achieved by forming a standing light wave from two counterpropagating laser beams, of which one is linearly polarized and the other is circularly polarized. We then predict the existence of a spin-dependent diffraction effect, if a beam of electrons crosses the standing light wave with a momentum of about $1 m c$ along the polarization direction of the linearly polarized laser beam, where $m$ is the electron restmass and $c$ the vacuum speed of light.

The article is organized as follows. In section \ref{sec:conceptual_remarks}, we introduce and explain the parameters of our laser-electron scattering scenario. In section \ref{sec:theory_description} we define the mathematical framework for the description of the spin-dependent electron diffraction effect and discuss the outcome of an analytic solution in terms of time-dependent perturbation theory. We support these considerations with a relativistic quantum simulation in section \ref{sec:spin_dependent_electron_diffraction_simulation}. After having demonstrated the possibility of this type of two-photon spin-dependent electron diffraction in the Bragg regime, we consider the possibility of an experimental implementation of the effect at the Shanghai High Repetition Rate XFEL and Extreme Light Facility (SHINE) in section \ref{sec:experimental_considerations}. In the final outlook (section \ref{sec:conclusion_and_outlook}) we compare our new spin-dependent interaction scheme with other proposals for spin-dependent electron diffraction in the literature. In the appendix, we discuss the perturbative solution of the electron in the standing light wave (appendix \ref{sec:dirac_perturbation_theory}), a Taylor expansion of the analytic spin-dependent electron scattering formula (appendix \ref{sec:taylor_expansion_compton_tensor}), a perturbative solution for an interacting, quantized electron-photon system, from which a relation to Compton scattering is established (appendix \ref{sec:qed_perturbation_theory}) and expressions of the spin propagation matrix on the tilted spinor basis, which is used in this article (appendix \ref{sec:tilted_spin_propagation}).

\section{Conceptual remarks\label{sec:conceptual_remarks}}

As mentioned in the introduction, we want to demonstrate the discussed spin effect with a parameter setup which corresponds to the scenario in reference \cite{ahrens_2017_spin_non_conservation}. Accordingly, we consider electron diffraction at a monochromatic, standing light wave along the $x$-axis
\begin{multline}
A_\mu(\vec x,t) = \frac{1}{2} \left( a_\mu e^{-i k_l\cdot x} + a_\mu^* e^{i k_l\cdot x} \phantom{e^{i k'_l\cdot x}}\right. \\
\left.+ a_\mu^\prime e^{-i k'_l\cdot x} + a_\mu^{\prime*} e^{i k'_l\cdot x}\right)\,.\label{eq:laser_field}
\end{multline}
In Eq. \eqref{eq:laser_field} we have introduced the two momentum four-vectors of the two counterpropagating laser beams
\begin{equation}
 k^{\mu}_l=(k_l,\vec k_l)\,,\qquad k^{\prime\mu}_l=(k_l,-\vec k_l)\,,\label{eq:laser_photon_momenta}
\end{equation}
with wave vector $\vec k_l = k_l \vec e_x$, laser wave number $k_l$ and the two polarizations $a_\mu$ and $a_\mu^\prime$ of the left and right propagating laser beam. Throughout the paper, except the experimental section \ref{sec:experimental_considerations}, we set $c=\hbar=1$, in a Gaussian unit system, such that laser frequency $\omega$ equals the laser wave number $k_l$. The dot between the four-vectors symbolizes a four-vector contraction $k_l \cdot x = k_l^\mu x_\mu$ according to Einstein's sum convention
\begin{equation}
 a_\mu b^\mu = \sum_\mu a_\mu b^\mu\,,
\end{equation}
with space-time metric $g_{\mu\nu} = \textrm{diag}(1,-1,-1,-1)$. Also, we use the symbol $*$ for denoting complex conjugation. The right and left propagating beam is linearly and circularly polarized, respectively and described by the corresponding polarization four-vectors
\begin{equation}
a = (0,0,0,\mathfrak{A})^\T\,,\quad a' = (0,0,\mathfrak{A}',i \mathfrak{A}')^\T/\sqrt{2}\,,\label{eq:special_polarization_A}
\end{equation}
where $\mathfrak{A}$ and $\mathfrak{A}'$ are the field amplitudes of the lasers' vector potentials and T denotes transposition. The electron has the initial momentum
\begin{equation}
 \tilde {\vec p}_i = -\vec k_l + m \vec e_z \label{eq:initial_electron_momentum}
\end{equation}
and we consider the two $45^\circ$ tilted spin states
\begin{equation}
 s^\searrow =
\begin{pmatrix}
 \cos 11 \pi/8\\
 \sin 11 \pi/8
\end{pmatrix}\,,\qquad
 s^\nwarrow =
\begin{pmatrix}
 \cos 15 \pi/8\\
 \sin 15 \pi/8
\end{pmatrix}\label{eq:tilted_spin_states}
\end{equation}
as initial electron spin configurations in this work.

In this following paragraph, we want to give a rough explanation of why the parameters \eqref{eq:special_polarization_A}, \eqref{eq:initial_electron_momentum} and \eqref{eq:tilted_spin_states} are taken as they are. Though spin-dependent terms may appear in electron-laser interactions, they are usually dominated by a spin-\emph{independent} term, which can be associated with the ponderomotive potential of the laser beam. Thus, spin-dynamics are usually superimposed by pronounced, spin-independent Rabi oscillations \cite{ahrens_2017_spin_filter} which potentially average out the spin effect. However, it is possible that the dominant contribution from the ponderomotive potential can cancel away, for certain configurations of the electron momentum and the laser polarization \cite{ahrens_bauke_2013_relativistic_KDE}. It seems that there is a continuum of parameters in parameter space (transverse electron momentum and laser polarizations), for which the spin-preserving terms are suppressed. The discussion about the structure of such a parameter space is beyond the scope of this work, but an investigation which shows a continuous variation of parameters, for which experimentally suitable spin dynamics may appear, is under study \cite{wang_yang_paper}. Regarding this article, the related parameters in the follow-up study \cite{ahrens_2017_spin_non_conservation} of reference \cite{ahrens_bauke_2013_relativistic_KDE} were constructed according to systematic reasoning. Since this specific spin effect is investigated with particular care in reference  \cite{ahrens_2017_spin_non_conservation}, we prefer to use the parameters in Eqs. \eqref{eq:special_polarization_A}, \eqref{eq:initial_electron_momentum} and \eqref{eq:tilted_spin_states} over other possible choices of parameters.

In this context, we would like to point out that reference \cite{ahrens_2017_spin_non_conservation} discusses spin dynamics in Compton scattering, whereas in this article, the in- and outgoing photon of the scattering process is substituted by two counter-propagating laser beams. This means that we describe the spin-dependent electron quantum dynamics in an external classical field of the counter-propagating laser beam background in terms of the Furry picture \cite{Furry_1951_Furry_picture,Landau_Lifshitz_1982_Quantum_Electrodynamics,Fradkin_Gitman_Shvartsman_1991_Quantum_Electrodynamics_with_Unstable_Vacuum}. In the limit of low field amplitudes however, where processes linear in the external field amplitudes are of relevance only, both described scenarios (Compton scattering and electron diffraction) have identical scattering amplitudes. Note, that this association of Compton scattering for electrons in low external fields was already pointed out by Ritus, where the Klein-Nishina formula and also the Breit-Wheeler formula were recovered in the low-field limit of an electron (described by the Dirac equation) in a plane wave field \cite{Ritus_1985_strong_field_solution_discussion}.

\begin{figure}%
	\includegraphics[width=0.47\textwidth]{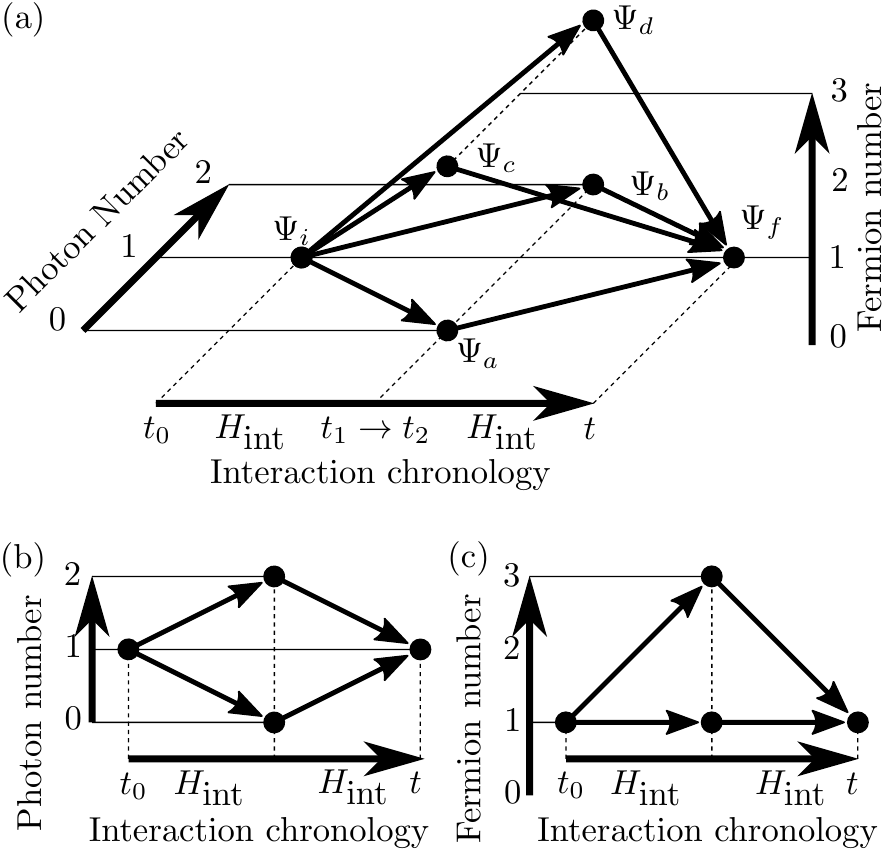}
	\caption{\label{fig:electron_photon_states}%
		Particle fluctuations of an interacting photon and electron. (a) When considering the lowest order interaction of an electron with a photon in a quantized photon electron description, one encounters the four different particle fluctuation configurations (quantum states in Eqs. \eqref{eq:intermediate_states}). In the photonic sector in panel (b) one either has absorption and then emission of the incoming and outgoing photon, or one has first emission and then absorption of the outgoing and incoming photon. In the electronic sector in panel (c) one encounters two quantum trajectories, in which either the electron propagates from its initial to its final state or in which the initial electron is accompanied by a virtual electron-positron pair, which then annihilates with the pair's anti-particle, with the final electron state remaining. The pairwise combination of the two times two processes in the panels (b) and (c) gives the four combinations in panel (a). One can associate these four quantum paths with Feynman graphs, as illustrated in figure \ref{fig:feynman_diagram_split_up}. Note however, that in contrast to the graphical conventions in Fig. \ref{fig:feynman_diagram_split_up}, which correspond to Feynman graphs, the roles of vertices and arrows are interchanged in the graphical representation in this figure: The big black dot corresponds to the free propagation of the quantum state, whereas the arrows indicate a change of the quantum state which is caused by the interaction $H_{\textrm{int}}$. See appendix \ref{sec:qed_perturbation_theory} for more information.}%
\end{figure}

\begin{figure}%
	\includegraphics[width=0.45\textwidth]{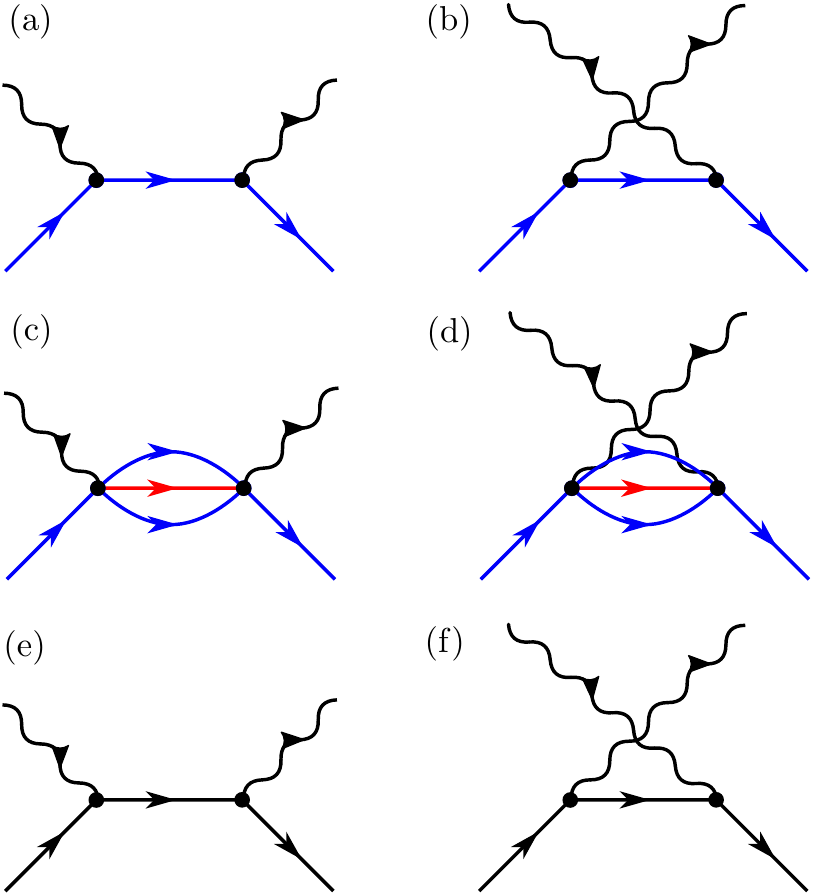}
	\caption{\label{fig:feynman_diagram_split_up}%
		(Color online)
		Association of Feynman graphs with a corresponding split-up version of the electron-propagator. One can show that the four quantum paths in Fig. \ref{fig:electron_photon_states} with $\Psi_a$, $\Psi_b$, $\Psi_c$ and $\Psi_d$ sketched in panels (a), (b), (c) and (d), respectively, can be combined into the Feynman graphs (e) and (f). More precisely, panel (e) is composed of the processes in panels (a) and (c) and panel (f) is composed of the processes in panels (b) and (d). The corresponding mathematical identification is carried out in appendix \ref{sec:compton_tensor_identification}.
	}%
\end{figure}

The match of Compton scattering and electron diffraction for low fields can be mathematically justified by showing that the perturbative solution of electron quantum dynamics in an external laser field (see appendix \ref{sec:dirac_perturbation_theory}) can be reformulated into perturbative scattering dynamics of one electron and one photon in the context of an interacting many particle electron-photon quantum system. This solution of the single electron-photon interaction can, in turn, be cast into the form of the Compton scattering formula (explained in appendix \ref{sec:compton_tensor_identification}). We have sketched this lowest order electron-photon interaction process in context of virtual particle fluctuations during the interaction in Fig. \ref{fig:electron_photon_states}. The appearing, four different intermediate particle states, denoted by $\Psi_a$, $\Psi_b$, $\Psi_c$ and $\Psi_d$ can be associated with the four diagrams (a), (b), (c) and (d) in Fig. \ref{fig:feynman_diagram_split_up}, which can be further summed up to give the two, vertex exchanged contributions of the Feynman graphs (e) and (f) of Compton scattering.
%
The solutions of time-dependent perturbation theory of a quantized electron-photon system as in Fig. \ref{fig:electron_photon_states} are known as old-fashioned perturbation theory (see literature \cite{Halzen_Martin_1984_quarks_and_leptons,Weinberg_1995_quantum_theory_of_fields}). Beyond the qualitative picture which is discussed in Figs. \ref{fig:electron_photon_states} and \ref{fig:feynman_diagram_split_up}, we also give a specific calculation in our article, where both perturbative derivations of the processes can be found in the appendices \ref{sec:dirac_perturbation_theory} and \ref{sec:qed_perturbation_theory}.

Note, that for the computation of the electron dynamics in the two laser beams, we have chosen the monochromatic standing light wave configuration \eqref{eq:laser_field} with laser photon momenta \eqref{eq:laser_photon_momenta}, because such an arrangement seems to be more common and is also more suitable for a numerical computation. In general, one could also consider bi-chromatic dynamics or dynamics with non-parallel laser beams. Such a general scenario could however then be related by a Lorentz transformation to our described scenario. In the context of the chosen laser photon momenta \eqref{eq:laser_photon_momenta} and the initial electron momentum \eqref{eq:initial_electron_momentum}, a non-trivial interaction with each of the laser beams results in the final electron momentum
\begin{equation}
\tilde {\vec p}_f = \vec k_l + m \vec e_z\,, \label{eq:final_electron_momentum}
\end{equation}
as implied by momentum conservation. The longitudinal $x$-component of the initial and final electron momentum is thereby chosen such that also energy is conserved for the electron and photon which constitute to the scattering process in a classical picture.

\section{Theoretical description\label{sec:theory_description}}

The electron quantum dynamics is computed by making a plane wave decomposition of the electron wave function
\begin{equation}
\psi(\vec x,t) = \sum_{n,s} \left( c_n^s(t) u^s_{\vec k_n} e^{-i \vec k_n \cdot \vec x} + d_n^s(t) v^s_{-\vec k_n} e^{-i \vec k_n \cdot \vec x} \right)\,.\label{eq:electron_wavefunction}
\end{equation}
The approach allows for the transfer of multiple photon momenta $\vec k_n = \vec p_i + n \vec k_L$, with the partial wave's complex amplitudes $c_n^s(t)$, $d_n^s(t)$ for positive and negative solutions, respectively. The positive and negative solutions of the free Dirac equation are the bi-spinors
\begin{subequations}%
\begin{align}%
 u^s_{\vec k} &=
 \sqrt{\frac{m}{\mathcal{E}_{\vec k}}}
 \sqrt{\frac{\mathcal{E}_{\vec k} + m}{2 m}}
 \begin{pmatrix}
  \chi^s \\
  \frac{\vec \sigma \cdot \vec k}{\mathcal{E}_{\vec k} + m}\, \chi^s
 \end{pmatrix}\\
 v^s_{\vec k} &=
 \sqrt{\frac{m}{\mathcal{E}_{\vec k}}}
 \sqrt{\frac{\mathcal{E}_{\vec k} + m}{2 m}}
 \begin{pmatrix}
  \frac{\vec \sigma \cdot \vec k}{\mathcal{E}_{\vec k} + m} \chi^s\\
  \chi^s
 \end{pmatrix}\,,\label{eq:bi-spinor_positron}
\end{align}\label{eq:bi-spinors}%
\end{subequations}%
where $s$ denotes the spin of each wave. The $\vec p_i$ is the initial electron momentum, whose transverse component can differ from Eq. \eqref{eq:initial_electron_momentum} at the stage of derivation and $\mathcal{E}_{\vec k} = \sqrt{m^2 + \vec k^2}$ is the relativistic energy momentum relation of the electron. The vector $\vec \sigma$ is the vector $(\sigma_1,\sigma_2,\sigma_3)^\textrm{T}$ of the Pauli matrices
\begin{equation}
 \sigma_x=
\begin{pmatrix}
 0 & 1 \\ 1 & 0
\end{pmatrix}\,,\ 
 \sigma_y=
\begin{pmatrix}
 0 & -i \\ i & 0
\end{pmatrix}\,,\ 
 \sigma_z=
\begin{pmatrix}
 1 & 0 \\ 0 & -1
\end{pmatrix}\,,
\end{equation}
where we use the indices $\{1,2,3\}$ and $\{x,y,z\}$ interchangeably for indexing Pauli matrices in this article. The dot between 3 component spacial vectors in Eq. \eqref{eq:electron_wavefunction} is denoting the inner product in Euclidean space $\vec k_{n} \cdot \vec x = \sum_a \vec k_{n,a} x_a$. The two component objects $\chi^s = (\chi_1^s,\chi_2^s)^\textrm{T}$ denote spinors. Note that in Refs. \cite{ahrens_bauke_2013_relativistic_KDE,ahrens_2017_spin_filter} the spinors $v^{s}_{\vec k}$  (see Eq. \eqref{eq:bi-spinor_positron}) have been introduced with opposite momentum $\vec k$. We also point out, that we have absorbed the phase space factor $(m/\mathcal{E}_{\vec k})^{1/2}$ from the Compton cross section formula into the normalization of the spinor definition \eqref{eq:bi-spinors}. Furthermore, we mention that the form of the wave function's plane wave expansion in Eq. \eqref{eq:electron_wavefunction} is an implication from the standing light wave \eqref{eq:laser_field}.

The time evolution of the wave function \eqref{eq:electron_wavefunction} in terms of its expansion coefficients can be formally written as
\begin{align}
 c^s_n(t) &= \sum_{a,s'}\left[U_{n,a}^{+,s;+,s'}(t,0) c_a^{s'}(0) + U_{n,a}^{+,s;-,s'}(t,0) d_a^{s'}(0)\right]\nonumber \\
 d^s_n(t) &= \sum_{a,s'}\left[U_{n,a}^{-,s;+,s'}(t,0) c_a^{s'}(0) + U_{n,a}^{-,s;-,s'}(t,0) d_a^{s'}(0)\right]\,.\label{eq:wave_function_propagation}
\end{align}
A perturbative expression of the propagation functions $U_{n,a}^{\gamma,s;\gamma',s'}(t,0)$ can be provided by transforming the Dirac equation
\begin{multline}
i \dot \psi(\vec x,t) = \left[ \left( -i \vec \nabla - \textrm{e} \vec A(\vec x,t) \right)\cdot \vec \alpha + m \beta \phantom{A^0}\right.\\
+ \left. \textrm{e} A^0(\vec x,t)\right] \psi(\vec x,t)\label{eq:dirac_equation}
\end{multline}
into momentum space and applying second order time-dependent perturbation theory to the resulting equations of motion. In Eq. \eqref{eq:dirac_equation}, the $\vec \alpha$ and $\beta$ are the Dirac matrices
\begin{equation}
 \alpha_i=
\begin{pmatrix}
 0 & \sigma_i \\ \sigma_i & 0
\end{pmatrix}\,,\quad
 \beta=
\begin{pmatrix}
 \mathds{1} & 0 \\ 0 & \mathds{-1}
\end{pmatrix}
\,,
\end{equation}
where $\mathds{1}$ is the $2 \times 2$ identity matrix. The elementary electric charge $\textrm{e}$ simplifies into the square root of the fine structure constant $\textrm{e}=\sqrt{\alpha}$ in our chosen unit system. The dot on top of the wave function $\psi(\vec x,t)$ in Eq. \eqref{eq:dirac_equation} denotes its time derivative $\dot \psi(\vec x,t) = \partial \psi(\vec x,t)/\partial t$. The procedure of rewriting and perturbatively solving the Dirac equation in a momentum space description is similar to corresponding procedures in references \cite{ahrens_bauke_2013_relativistic_KDE,bauke_ahrens_2014_spin_precession_2}. Therefore, we have shifted the details of the calculation to appendix \ref{sec:dirac_perturbation_theory} and focus on the physics description here.

The perturbative solution of the spin dependent quantum state propagation of the initial electron spin state $c_0(0)$ to the final electron spin state $c_2(0)$ is proportional to the matrix
\begin{equation}
 M_s =\frac{1}{\sqrt{8}}
\begin{pmatrix}
 -1 & -1 - \sqrt{2} \\
 -1 + \sqrt{2} & 1
\end{pmatrix}
= s^\nwarrow \cdot s^{\searrow\dagger}\label{eq:tilted_spin_filter}
\end{equation}
for our chosen parameters of the photon polarization \eqref{eq:special_polarization_A} and the initial electron momentum \eqref{eq:initial_electron_momentum}. We take this spin-propagation matrix from the Taylor expansion of the spin propagation matrix $M$ in appendix \ref{sec:taylor_expansion_compton_tensor}. In this context we point out that we desire that spin preserving terms (terms proportional to $\mathds{1}$) cancel in the electron spin dynamics, as mentioned above. This is approximately the case for the transverse momenta of $\tilde {\vec p}_i$ in Eq. \eqref{eq:initial_electron_momentum}. However, small corrections remain, such that we choose
\begin{equation}
\left(\tilde {\vec p}_i\right)_3 \approx \left(1 + 1.34... \cdot 10^{-4} \right) m\label{eq:z-component_tuning}
\end{equation}
for the $z$-component of the electron momentum in our numerical simulation with the selected laser photon energy of $k_l = 13\,\textrm{keV} = 0.025... m$. Nevertheless, the corrections to $\tilde {\vec p}_i$ are more than two orders of magnitude smaller than the photon momentum $k_l$ itself, such that the correction \eqref{eq:z-component_tuning} has no strong influence on the physics which is discussed in this work.

The right-hand side of Eq. \eqref{eq:tilted_spin_filter} shows an outer product representation of $M_s$, created from the pair of two-component spinors $s^\nwarrow$ and $s^\searrow$. This can be seen from their expressions
\begin{subequations}%
\begin{align}%
 s^{\searrow} &=
 \begin{pmatrix}
  1 - \sqrt{2} \\ -1
 \end{pmatrix}
\sqrt{2 (2 - \sqrt{2})}^{-1}\\
 s^{\nwarrow} &=
 \begin{pmatrix}
  1 + \sqrt{2} \\ -1
 \end{pmatrix}
\sqrt{2 (2 + \sqrt{2})}^{-1}\,,
\end{align}\label{eq:tilted_spin_states_alternative_definition}%
\end{subequations}%
which are equivalent to the definitions \eqref{eq:tilted_spin_states} and in consistence with the convention in reference \cite{ahrens_2017_spin_non_conservation}. From the outer product representation of the matrix $M_s$ at the right-hand side of Eq. \eqref{eq:tilted_spin_filter} one immediately obtains
\begin{subequations}%
\begin{align}%
 \braket{s^\searrow|M_s|s^\searrow} &= 0 & \braket{s^\nwarrow|M_s|s^\searrow} &= 1\\
 \braket{s^\searrow|M_s|s^\nwarrow} &= 0 & \braket{s^\nwarrow|M_s|s^\nwarrow} &= 0\,,\label{eq:spin-dependent_diffraction_nw_initial}
\end{align}\label{eq:spin-dependent_diffraction}%
\end{subequations}%
which is consistent with the corresponding scenario in Compton scattering \cite{ahrens_2017_spin_non_conservation}, where a $s^\searrow$ polarized electron is scattered at a vertically polarized photon into a left circularly polarized photon and a $s^\nwarrow$ polarized electron. The opposite scenario, where a $s^\nwarrow$ polarized electron is scattered into a right circularly polarized photon and a $s^\searrow$ polarized electron (see reference \cite{ahrens_2017_spin_non_conservation}) is considered to be overruled by the effect of induced emission into a left circularly polarized photon for the case of coherent electron diffraction with the laser polarization \eqref{eq:special_polarization_A} in the Kapitza-Dirac scattering.

\section{Numerical solution of the spin-dependent quantum dynamics\label{sec:spin_dependent_electron_diffraction_simulation}}

\begin{figure}%
	\includegraphics[width=0.475\textwidth]{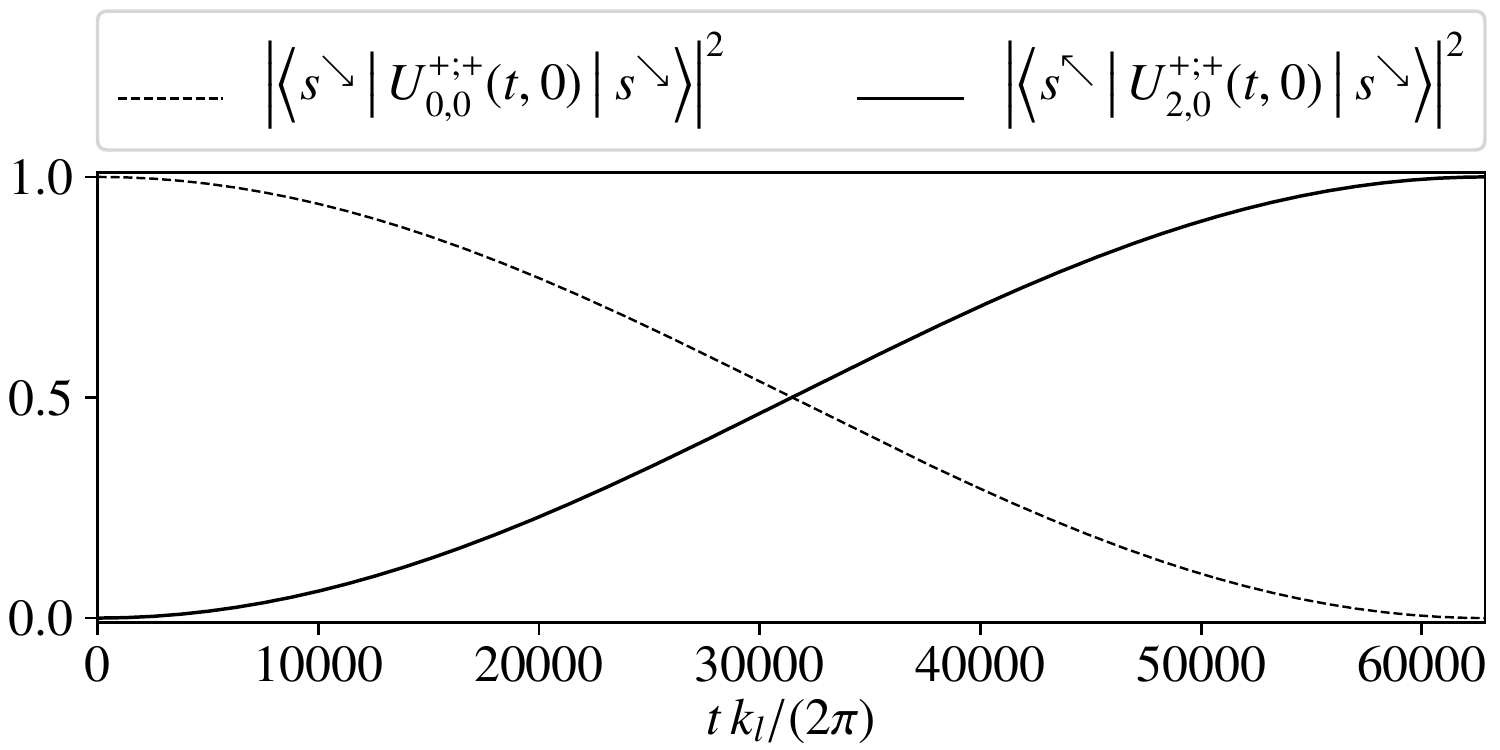}
	\caption{\label{fig:electron_dynamics}%
		Simulated spin-dependent electron diffraction effect. The panel shows the two non-vanishing matrix elements of the numerical solution of the Dirac equation in momentum space \eqref{eq:dirac_equation_in_momentum_space}, represented by the quantum state propagation \eqref{eq:wave_function_propagation}. The simulation is carried out with the field amplitudes $\mathfrak{A}$ and $\mathfrak{A'}$ and laser photon energy $k_l$ as specified in Eq. \eqref{eq:simulation_parameters}, where the initial electron momentum is $\tilde {\vec p}_i$ (see Eq.  \eqref{eq:initial_electron_momentum}). One can see that the system follows the simple Rabi model \eqref{eq:rabi_model}, from an electron with initial momentum $k_0$ and spin state $s^\searrow$ into an electron with final momentum $k_2$ and spin state $s^\nwarrow$. For the opposite initial spin configuration $s^\nwarrow$ we find that no such Rabi oscillations occur, which demonstrates a theoretically perfect spin filtering and spin polarization effect.
	}%
\end{figure}

We support the above considerations by performing numerical simulations of the one-particle Dirac equation in momentum space \eqref{eq:dirac_equation_in_momentum_space} (ie. a numerical solution of the propagation equation \eqref{eq:wave_function_propagation}), of an electron in a standing wave of light in Fig. \ref{fig:electron_dynamics}. Such a procedure is similar to the numerical simulations shown in references \cite{ahrens_bauke_2012_spin-kde,ahrens_bauke_2013_relativistic_KDE,bauke_ahrens_2014_spin_precession_1,bauke_ahrens_2014_spin_precession_2,erhard_bauke_2015_spin,dellweg_awwad_mueller_2016_spin-dynamics_bichromatic_laser_fields,dellweg_mueller_extended_KDE_calculations,ahrens_2017_spin_filter,ebadati_2018_four_photon_KDE}. In the simulation, the standing wave of light \eqref{eq:laser_field} has the polarization \eqref{eq:special_polarization_A}. The standing wave's field amplitude is smoothly ramped up and down for the duration of five laser periods at the beginning and the end of the simulation by a $\sin^2$ temporal envelope, as done in the references \cite{ahrens_bauke_2012_spin-kde,ahrens_bauke_2013_relativistic_KDE,bauke_ahrens_2014_spin_precession_1,bauke_ahrens_2014_spin_precession_2,erhard_bauke_2015_spin,dellweg_awwad_mueller_2016_spin-dynamics_bichromatic_laser_fields,dellweg_mueller_extended_KDE_calculations,ahrens_2017_spin_filter,ebadati_2018_four_photon_KDE}. 

In the numerical simulation, we see no diffraction dynamics for an electron with initial $s^\nwarrow$ spin configuration, ie. we have
\begin{subequations}%
\begin{align}%
 \left|\Braket{s^\nwarrow | U_{0,0}^{+;+}(t,0) | s^\nwarrow }\right|^2 &\approx 1\\
 \left|\Braket{s^\searrow | U_{0,0}^{+;+}(t,0) | s^\nwarrow }\right|^2 &\approx 0\\
 \left|\Braket{s^\nwarrow | U_{2,0}^{+;+}(t,0) | s^\nwarrow }\right|^2 &\approx 0\\
 \left|\Braket{s^\searrow | U_{2,0}^{+;+}(t,0) | s^\nwarrow }\right|^2 &\approx 0\,.
\end{align}%
\end{subequations}%
This is consistent to our analytic considerations from perturbation theory, see Eq. \eqref{eq:spin-dependent_diffraction_nw_initial}. For this reason we only show the projection on the diffracted spin state%
\begin{subequations}%
\begin{equation}%
 \left|\Braket{s^\nwarrow | U_{2,0}^{+;+}(t,0) | s^\searrow }\right|^2
\end{equation}%
and the initial quantum state%
\begin{equation}%
 \left|\Braket{s^\searrow | U_{0,0}^{+;+}(t,0) | s^\searrow }\right|^2
\end{equation}\label{eq:non_vanishing_matrix_elements}%
\end{subequations}%
of the numerical solution of the propagation $U^{+,+}_{a,b}(t,0)$ in Fig. \ref{fig:electron_dynamics}. In fact, these are the only non-negligible contributions of the time evolution. Equivalently, one can say that
both expressions \eqref{eq:non_vanishing_matrix_elements} sum up to approximately 1, such that the unitarity of the Dirac equation implies that any other excitations are negligibly small.

Note, that the half period Rabi cycle, which is shown in Fig. \ref{fig:electron_dynamics} lasts for $6.29\times 10^4$ optical cycles of the laser field, corresponding to the Rabi frequency
\begin{equation}
\Omega_R=2.02\times 10^{-7}\,m\label{eq:rabi_frequency}
\end{equation}
in the effective Rabi model
\begin{equation}
 \left|\Braket{s^\nwarrow | U_{2,0}^{+;+}(t,0) | s^\searrow }\right|^2 = \sin\left(\frac{\Omega_R t}{2}\right)^2\,.\label{eq:rabi_model}
\end{equation}
This is consistent with the approximate equation for the matrix element
\begin{equation}
 \left|\Braket{s^\nwarrow | U_{2,0}^{+;+}(t,0) | s^\searrow }\right|^2 \approx \left( \frac{\textrm{e}\mathfrak{A}\textrm{e}\mathfrak{A}' k_l t}{8 m^2\sqrt{2}} \right)^2\label{eq:diffraction_probability}
\end{equation}
of the perturbative solution of the Dirac equation in Eq. \eqref{eq:perturbative_propagator_taylor}. In this context, we assume that the left-hand side of Eq. \eqref{eq:diffraction_probability} can be identified with the analytic short time approximation of the Rabi model \eqref{eq:rabi_model}
\begin{equation}
 \left|\Braket{s^\nwarrow | U_{2,0}^{+;+}(t,0) | s^\searrow }\right|^2 \approx \left(\frac{\Omega_R t}{2}\right)^2\,,
\end{equation}
where we have set the parameters
\begin{subequations}%
\begin{align}%
 \textrm{e} \mathfrak{A}/m = \textrm{e} \mathfrak{A'}/m &= 4.74 \times 10^{-2} \label{eq:simulation_parameters_field_amplitude}\\
 k_l/m&=2.54 \times 10^{-2} \label{eq:simulation_parameters_photon_energy}
\end{align}\label{eq:simulation_parameters}%
\end{subequations}%
in our numerical simulation. We choose the photon energy to be 13\,keV, corresponding to the value of $k_l$ in Eq. \eqref{eq:simulation_parameters_photon_energy} for the simulation. Similarly, we have set the simulation's laser field amplitude $\mathfrak{A}$ and $\mathfrak{A'}$ in \eqref{eq:simulation_parameters_field_amplitude}, such that a half Rabi period will last exactly 20\,fs. This value corresponds to the value of the Rabi frequency \eqref{eq:rabi_frequency}.

The actual numerical implementation was carried out in the  basis of the states $c^s_n$ and $d^s_n$ with spin up and spin down $s \in \{\uparrow,\downarrow\}$, where the matrix elements with respect to the spin states $s^\searrow$ and $s^\nwarrow$ of the numerical propagation $U^{+,+}_{a,b}(t,0)$ are given explicitly in appendix \eqref{sec:tilted_spin_propagation}. Note that the transition amplitudes $U_{n,0}(t,0)$ of higher momentum states $|n|$ are dropping off exponentially for the chosen parameters in Eq. \eqref{eq:simulation_parameters}, such that we have truncated the higher modes in the numerical solutions at $|n|=12$, similar to the procedure in references \cite{ahrens_bauke_2012_spin-kde,ahrens_bauke_2013_relativistic_KDE,bauke_ahrens_2014_spin_precession_1,bauke_ahrens_2014_spin_precession_2,erhard_bauke_2015_spin,dellweg_awwad_mueller_2016_spin-dynamics_bichromatic_laser_fields,dellweg_mueller_extended_KDE_calculations,ahrens_2017_spin_filter,ebadati_2018_four_photon_KDE}.

We want to point out that Eqs. \eqref{eq:spin-dependent_diffraction} demonstrate a spin-dependent diffraction effect: While the initial spin configuration $s^\searrow$ is diffracted into a $s^\nwarrow$ configuration, an initial spin $s^\nwarrow$ is not diffracted at all! Thus, electrons are filtered according to their initial spin orientation. Also, the outer product in \eqref{eq:tilted_spin_filter} implies that whatever electron spin is diffracted, the final electron spin will always be $s^\nwarrow$. This also demonstrates that the electron spin can be polarized by the diffraction mechanism. These two properties (filtering and polarization of the electron spin) are the same characterizations which we have already pointed out in our previous work \cite{ahrens_2017_spin_filter}, where a two-photon spin-dependent diffraction is presented which has similar properties as in this work. However, the spin-dependent diffraction effect in reference \cite{ahrens_2017_spin_filter} only appears after multiple Rabi cycles, whereas the spin-dependent diffraction in our current work appears already with the rise of the transition's oscillation in the form of a Bragg peak, which appears to be more suitable for the experimental implementation.

We also want to point out that the spin-dependent propagation matrix in Eq. \eqref{eq:tilted_spin_filter} is a generalization of our statement in reference \cite{ahrens_2017_spin_filter}, in which a spin-dependent quantum state propagation has been identified to be proportional to a projection matrix. In contrast, Eq. \eqref{eq:tilted_spin_filter} demonstrates explicitly that even a projection is not the most general characterization for spin-dependent dynamics. A general and specific criterion for spin-dependent diffraction dynamics might be non-trivial and be a subject of future investigations.

\section{Experimental implementation considerations\label{sec:experimental_considerations}}

\begin{figure}%
  \includegraphics[width=0.47\textwidth]{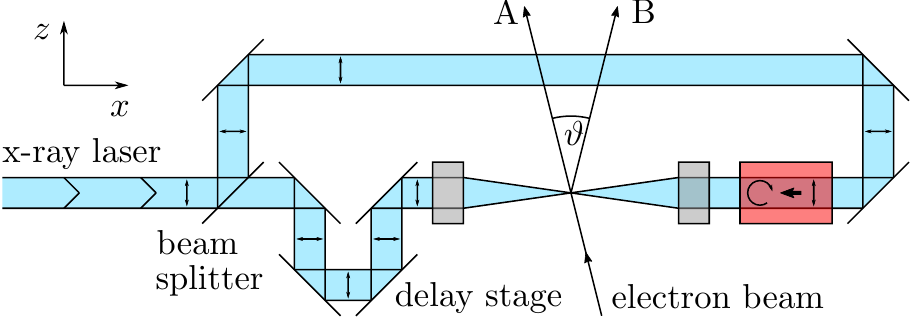}
  \caption{\label{fig:experimental_setup}%
(Color online) Possible experimental setup for establishing spin-dependent electron diffraction based on two photon Kapitza-Dirac scattering. A linearly polarized X-ray laser with 13\,keV photon energy is entering from the left. A part of the beam is transmitted through the beam splitter and the reflected part of the X-ray laser beam is guided to approach the beam focus from the opposite direction. A phase retarder (red box with an opened arrow circle) is converting the linear X-ray polarization into circular polarization. All X-ray optics are chosen such that the two counterpropagating laser beams are reaching the beam focus with equal intensity and at equal time. For the setup, spin-dependent diffraction is expected to be observed for initially spin-polarized electrons, which approach the laser focal spot with a kinetic energy of 212\,keV at an inclination angle $90^\circ - \vartheta/2$ to the beam propagation direction (see main text for details). The small, grey, rectangular boxes are symbolizing beam focussing optics.}%
\end{figure}

We want to discuss a possible experimental implementation of the spin-dependent laser electron interaction, according to the setup in Fig. \ref{fig:experimental_setup}. In this example, the source of the X-ray laser beams is assumed to be the Shanghai High Repetition Rate XFEL and Extreme Light Facility (SHINE), which is currently under construction \cite{Shen_2018_vacuum_birefringence}. Within its design parameters, SHINE will provide 100\,GW laser pules at 13\,keV photon energy and with a pulse duration of 20\,fs. When the beam is focused to 100\,nm, the peak intensity reaches $1.2\times 10^{21}\textrm{W}/\textrm{cm}^2$. A coincident laser pulse overlap at the interaction point is achieved by reflecting the two beams as in the arrangement in Fig. \ref{fig:experimental_setup}. Circular polarization can be converted from the linear polarized laser beam by utilizing a phase retardation setup in X-ray diffraction \cite{Suzuki_2014_polarization_control_diamond_phase_retarder}. In this way, two coincident, counterpropagating, high intensity pulses can be established at the beam focus, with a linearly polarized beam from the left and a circularly polarized beam from the right. By assuming mirror reflectivities of $85\%$, a phase retarder transmittivity of $55\%$ and a beam splitter design with $34\%$ transmission and $56\%$ reflection \cite{Osaka_2013_x_ray_beam_splitter} one estimates an intensity of $1.2\times 10^{20}\,\textrm{W}/\textrm{cm}^2$ for the left and right beam at the laser focus spot. Eq. \eqref{eq:diffraction_probability} can be written in terms of SI
units as
\begin{equation}
\left|\Braket{s^\nwarrow | U_{2,0}^{+;+}(t,0) | s^\searrow }\right|^2 \approx \left( \frac{\alpha \lambda_c^2}{8 \pi \sqrt{2}} \frac{I_1^{1/2}I_2^{1/2} t}{c \hbar k_l} \right)^2 \,.
\end{equation}
Here, $\alpha$ is the fine-structure constant and $\lambda_c$ the Compton wavelength and $I_1$ and $I_2$ are the intensities of the left- and right propagating laser beams. Evaluated with the parameters above, one is expecting a probability of about $1.1\cdot 10^{-7}$ for an electron with a spin $\searrow$ orientation to be diffracted in the direction of beam $B$. Since we are discussing a spin-dependent diffraction scheme, electrons with spin $\nwarrow$ orientation will not be diffracted into beam $B$. For undergoing spin-dependent diffraction, the electrons have to have the specific momentum of $511\,\textrm{keV}/c$ along the $z$-axis, corresponding to a kinetic energy 212\,keV. When undergoing diffraction, the electron will pick up two longitudinal photon momenta of $13\,\textrm{keV}/c$ along the $x$-axis. Since the momentum change is longitudinal, one can relate this to a diffraction angle of $\vartheta=2.9^\circ$ from the scattering geometry. Spin polarized electron pulses with charges of 10\,fC are available \cite{Kuwahara_2012_polarized_electron_pulses} and with the temporal electron bunch width of 10\,ps, one expects 124 electrons to cross the beam focal spot in its 20\,fs duration. Therefore, with the SHINE aimed repetition rate of 1\,MHz we estimate a countrate of 13 electrons per second for the spin-dependent electron diffraction effect. Similar parameters for establishing the considered experimental configuration can also be reached at the LCLS in Stanford \cite{lcls-website,*Lutman_2018_high_power_xfel} and the European X-FEL in Hamburg \cite{x-fel-website,*Altarell:etal:2007:XFEL}.

\section{Discussion and Outlook\label{sec:conclusion_and_outlook}}

In this article we have discussed a spin-dependent Kapitza-Dirac diffraction effect, which can be implemented in the form of a Bragg scattering setup and which requires only the interaction with two of the standing light wave's photons. Open questions for the effect are the influence of the laser beam focus on the spin-dependent electron dynamics. Within this article, we have treated the laser beam and also the electron wave function as a discrete superposition of a finite number of plane waves, whereas a Gaussian beam and a Gaussian wave packet would model electron and laser more realistically. In this context the question arises, how a small longitudinal field component \cite{Salamin_2006_gaussian_beams}, which is implied by the laser beam focus, is influencing the spin dynamics. Also, the contribution of spontaneous emission of electro-magnetic radiation as compared to the induced emission into the laser beam is of relevance and can be computed \cite{mocken_2005_relativistic_radiation_emission}. The question on how the quantum state of the laser field is modified by the electron diffraction dynamics is also of relevance, because the Compton scattering version of the effect raises questions about the transfer of intrinsic angular momentum (spin) between the electron and the photon \cite{ahrens_2017_spin_non_conservation}.

There are two possible laser frequency regimes for the implementation of the effect, which are realistic in terms of available laser intensity for the experiment: The optical regime and the x-ray regime.

The optical regime has the advantage that the classical nonlinearity parameter $\xi=\textrm{e}\mathfrak{A}/m$ can reach values of 1 with comparably low effort, such that high photon number Kapitza-Dirac scattering, as for example discussed in references \cite{ahrens_bauke_2012_spin-kde,ahrens_bauke_2013_relativistic_KDE,McGregor_Batelaan_2015_two_color_spin,dellweg_awwad_mueller_2016_spin-dynamics_bichromatic_laser_fields,dellweg_mueller_2016_interferometric_spin-polarizer,dellweg_mueller_extended_KDE_calculations,ebadati_2018_four_photon_KDE,Kozak_2018_ponderomotive_KDE,ebadati_2019_n_photon_KDE} could be possible. Note, that the short-time diffraction probability and the transition's Rabi frequency are proportional to the field amplitude $\mathfrak{A}$ to the power of the number of interacting photons, implying that either $\xi$ should be close to 1 or the number of contributing photons should be as small as possible. Bi-chromatic setups \cite{dellweg_mueller_extended_KDE_calculations,ebadati_2018_four_photon_KDE} appear promising for the experiment due to potentially long laser-electron interaction times caused by low initial and final electron momenta. However, one challenge with optical systems would be the control of the transverse electron momentum and the laser polarization such that the effect does not smear out. A look on the matrix \eqref{eq:taylor_expansion_compton_tensor} of the polarization dependent spin dynamics for the electron in the laser beam tells that the electron momentum should be under control on the order of the photon momentum $k_l$. Also the laser polarization should be controlled on the accuracy level $k_l/m$, where we have $k_l \approx 10^{-6}m$ in the optical regime.

In the x-ray regime, on the other hand, this need of fine tuning would be only at the percent level. Here, one faces the challenge of providing field amplitudes, such that $\xi$ is close to one, which might be possible for the case of small beam foci. 
Therefore, for implementing a spin-dependent diffraction setup for x-rays, a lower order photon interaction Kapitza-Dirac effect would be beneficial. Two photon scattering would be the lowest possible configuration for Kapitza-Dirac-like scattering, since a one-photon interaction is not compatible with the conservation of energy and momentum. A two-photon setup from a previous investigation which only depends on a longitudinal electron momentum \cite{erhard_bauke_2015_spin,ahrens_2017_spin_filter} appears to be promising. However, for this scenario one faces the challenge that the spin oscillations are dependent on simultaneous Rabi oscillations with an enhanced frequency by the factor $m/k_l$, which also would imply the necessity of fine tuning. In contrast, the spin-dependent two photon effect which is discussed within this article is \emph{not} superimposed by a larger spin-preserving term in the electron spin propagation. Therefore, only the beginning of a Rabi cycle (ie. the Bragg peak) of the diffraction effect would have to be observed for seeing the spin-dependent electron-laser interaction. For this reason, the
spin-dependent electron diffraction effect as discussed in section \ref{sec:experimental_considerations} appears to be suitable for implementing spin-dependent electron diffraction in standing light waves.

\begin{acknowledgments}
S. A. thanks C. M\"uller and C.-P. Sun for discussions. This work has been supported by the National Science Foundation of China (Grant Nos. 11975155 and 1935008) and the Ministry of Science and Technology of the People’s Republic of China (Grant Nos. 2018YFA0404803 and 2016YFA0401102). T. \v{C}. gratefully acknowledges the support by the Institute for Basic Science in Korea (IBS-R024-D1).
\end{acknowledgments}

\appendix

\section{Perturbative solution of the Dirac equation in an external standing light wave\label{sec:dirac_perturbation_theory}}



In this appendix section, we carry out a perturbative electron spin dynamics calculation, which is used in section \ref{sec:theory_description}. As mentioned, according to a similar procedure in references \cite{ahrens_bauke_2013_relativistic_KDE,bauke_ahrens_2014_spin_precession_2,ahrens_2017_spin_filter}, the Dirac equation \eqref{eq:dirac_equation} can be rewritten into a momentum space description with respect to the wave function ansatz \eqref{eq:laser_field} by projecting the plane wave eigensolutions $u^s_{\vec k_n} e^{-i \vec k_n \cdot \vec x}$ and $v^s_{-\vec k_n} e^{-i \vec k_n \cdot \vec x}$ of the Dirac equation from the left. This results in the coupled system of differential equations
\begin{subequations}%
	\begin{align}%
	i \dot c_n^s(t) &= \phantom{-}\mathcal{E}_{\vec k_n} c_n^s(t) + \sum_{n',s'}\bigg[ V^{+,s;+,s'}_{n,n'}(t) c_{n'}^{s'}(t) \nonumber\\
	& \hspace{2.9 cm}+ V^{+,s;-,s'}_{n,n'}(t) d_{n'}^{s'}(t) \bigg] \\
	i \dot d_n^s(t) &= -\mathcal{E}_{\vec k_n} d_n^s(t) + \sum_{n',s'}\bigg[ V^{-,s;+,s'}_{n,n'}(t) c_{n'}^{s'}(t) \nonumber\\
	& \hspace{2.9 cm}+ V^{-,s;-,s'}_{n,n'}(t) d_{n'}^{s'}(t) \bigg]\,,
	\end{align}\label{eq:dirac_equation_in_momentum_space}%
\end{subequations}%
where the potential interaction functions $V^{\gamma,s;\gamma',s'}_{n,n'}(t)$ are related to the standing light wave's potential \eqref{eq:laser_field} by
\begin{multline}
V^{\gamma,s;\gamma',s'}_{n,n'}(t) = - \frac{\textrm{e}}{2} L_{n,n'}^{\gamma,s;\gamma',s';\mu} \\
\times\bigg[ \left( a_\mu e^{-i k_l t} + a_{\mu}^{\prime*} e^{i k_l t} \right) \delta_{n',n+1} \\
+ \left( a_{\mu}^{*} e^{i k_l t} + a_{\mu}^{\prime} e^{-i k_l t} \right)\delta_{n',n-1} \bigg]\,.
\end{multline}
Here, we have introduced the additional expressions
\begin{subequations}%
	\begin{align}%
	L_{n,n'}^{+,s;+,s';\mu} &= u^{s\dagger}_{\vec k_n} \gamma^0 \gamma^\mu u^{s'}_{\vec k_{n'}} \\
	L_{n,n'}^{+,s;-,s';\mu} &= u^{s\dagger}_{\vec k_n} \gamma^0 \gamma^\mu v^{s'}_{-\vec k_{n'}} \\
	L_{n,n'}^{-,s;+,s';\mu} &= v^{s\dagger}_{-\vec k_n} \gamma^0 \gamma^\mu u^{s'}_{\vec k_{n'}} \\
	L_{n,n'}^{-,s;-,s';\mu} &= v^{s\dagger}_{-\vec k_n} \gamma^0 \gamma^\mu v^{s'}_{-\vec k_{n'}}\,.
	\end{align}\label{eq:spinor_matrix_contractions}%
\end{subequations}%
as generalized spin- and polarization dependent coupling terms, where $\gamma^0 = \beta$ and $\gamma^i = \beta \alpha_i$ are the Dirac gamma matrices. The dagger symbol $\dagger$ denotes combined complex conjugation and transposition.

One can establish a second order perturbative approximation of the quantum state propagation \eqref{eq:wave_function_propagation} by (see for example \cite{ahrens_2012_phdthesis_KDE})
\begin{multline}
U(t,t_0) \approx \frac{1}{i^2} \int_{t_0}^t d t_2 \int_{t_0}^{t_2} d t_1 \\
\times U_0(t,t_2) V(t_2) U_0(t_2,t_1) V(t_1) U_0(t_1,t_0)\,,\label{eq:dirac_second_order_perturbation}
\end{multline}
where $U$ and $V$ are matrices with the matrix product
\begin{multline}
\left[U_0(t_2,t_1) V(t_1)\right]_{n,n^{\prime\prime}}^{\gamma,s;\gamma^{\prime\prime},s^{\prime\prime}} \\
= \sum_{n',\gamma',s'} U_{0;n,n'}^{\gamma,s;\gamma',s'}(t_2,t_1) V_{n',n^{\prime\prime}}^{\gamma',s';\gamma^{\prime\prime},s^{\prime\prime}}(t_1)\,.
\end{multline}
The perturbative propagator \eqref{eq:dirac_second_order_perturbation} makes use of the expressions $U_0(t,t_0)$, which denote the free propagation
\begin{subequations}%
\begin{align}%
U_{0;n,n'}^{+,s;+,s'}(t,t_0) &= \exp\left(-i\mathcal{E}_{\vec k_n}(t-t_0)\right) \delta_{n,n'} \delta_{s,s'}\\
U_{0;n,n'}^{-,s;-,s'}(t,t_0) &= \exp\left(\phantom{-}i\mathcal{E}_{\vec k_n}(t-t_0)\right) \delta_{n,n'} \delta_{s,s'}\\
U_{0;n,n'}^{+,s;-,s'}(t,t_0) &= U_{n,n'}^{-,s;+,s'}(t,t_0) = 0
\end{align}%
\end{subequations}%
for the momentum space expansion coefficients $c_n^s$ and $d_n^s$. In section \ref{sec:conceptual_remarks} and \ref{sec:theory_description} we have introduced the setup of the electron and the standing light wave, such that the electron with initial momentum $\vec k_0$ can be scattered into the final momentum state $\vec k_2$, such that energy is conserved for the electron and the interacting photons. Interaction terms which result in this final momentum $\vec k_2$ will grow linear in time in the perturbative expression \eqref{eq:dirac_second_order_perturbation} and can dominate other contributions. Such a linear growth leads to Rabi oscillations, if one would account for all higher perturbation orders, as implied by the unitary time evolution (see reference \cite{gush_gush_1971_higher_order_kapitza_dirac_scattering} and also Figure \ref{fig:electron_dynamics}). By accounting only for the mentioned resonant terms, we obtain
\begin{subequations}%
	\begin{align}%
	&U_{2,0}^{+,s';+,s}(t,t_0) \approx \frac{\textrm{e}^2 a_{\mu}^{\prime *} a_{\nu}}{4 i^2} \sum_{s^{\prime\prime}}\int_{t_0}^t d t_2 \int_{t_0}^{t_2} d t_1 \bigg\{ \\
	&\hspace{1.63 cm}
	L_{2,1}^{+,s';+,s^{\prime\prime};\mu} L_{1,0}^{+,s^{\prime\prime};+,s;\nu} \xi_a(t,t_2,t_1,t_0) \\
	&\hspace{1.2 cm}
	+ L_{2,1}^{+,s';+,s^{\prime\prime};\nu} L_{1,0}^{+,s^{\prime\prime};+,s;\mu}\xi_b(t,t_2,t_1,t_0) \\
	&\hspace{1.2 cm}
	+ L_{2,1}^{+,s';-,s^{\prime\prime};\nu} L_{1,0}^{-,s^{\prime\prime};+,s;\mu} \xi_c(t,t_2,t_1,t_0) \\
	&\hspace{1.2 cm}
	+ L_{2,1}^{+,s';-,s^{\prime\prime};\mu} L_{1,0}^{-,s^{\prime\prime};+,s;\nu}\xi_d(t,t_2,t_1,t_0)  \bigg\}\,,
	\end{align}\label{eq:unintegrated_perturbative_propagator_dirac_equation}%
\end{subequations}%
with the phases
\begin{align}
\xi_a &= \exp\left[ -i \mathcal{E}_{\vec k_2} t + i \left(\mathcal{E}_{\vec k_0} - \mathcal{E}_{\vec k_1} + k_l\right) (t_2 - t_1) + i \mathcal{E}_{\vec k_0} t_0 \right] \nonumber \\
\xi_b &= \exp\left[ -i \mathcal{E}_{\vec k_2} t + i \left(\mathcal{E}_{\vec k_0} - \mathcal{E}_{\vec k_1} - k_l\right) (t_2 - t_1) + i \mathcal{E}_{\vec k_0} t_0 \right] \nonumber \\
\xi_c &= \exp\left[ -i \mathcal{E}_{\vec k_2} t + i \left(\mathcal{E}_{\vec k_0} + \mathcal{E}_{\vec k_1} - k_l\right) (t_2 - t_1) + i \mathcal{E}_{\vec k_0} t_0 \right] \nonumber \\
\xi_d &= \exp\left[ -i \mathcal{E}_{\vec k_2} t + i \left(\mathcal{E}_{\vec k_0} + \mathcal{E}_{\vec k_1} + k_l\right) (t_2 - t_1) + i \mathcal{E}_{\vec k_0} t_0 \right]\,,\label{eq:phases_dirac_equation}
\end{align}
where the argument $(t,t_2,t_1,t_0)$ is left away at the left-hand side of Eqs. \eqref{eq:phases_dirac_equation}. We made use of $\mathcal{E}_{\vec k_2} = \mathcal{E}_{\vec k_0}$ in Eqs. \eqref{eq:phases_dirac_equation}, as implied by energy conservation. The phase terms
\begin{subequations}%
\begin{align}%
&+ i \left(\mathcal{E}_{\vec k_0} - \mathcal{E}_{\vec k_1} + k_l\right) (t_2 - t_1) \\
&+ i \left(\mathcal{E}_{\vec k_0} - \mathcal{E}_{\vec k_1} - k_l\right) (t_2 - t_1) \\
&+ i \left(\mathcal{E}_{\vec k_0} + \mathcal{E}_{\vec k_1} - k_l\right) (t_2 - t_1) \\
&+ i \left(\mathcal{E}_{\vec k_0} + \mathcal{E}_{\vec k_1} + k_l\right) (t_2 - t_1)
\end{align}%
\end{subequations}%
produce the mentioned, linear growth behavior in the integral over $t_2$, for the upper limit of the $t_1$ integration, such that we obtain
\begin{subequations}%
\begin{align}%
&U_{2,0}^{+,s';+,s}(t,t_0) \approx\\
&\hspace{0.3 cm}-i \frac{\textrm{e}^2 a_{\mu}^{\prime *} a_{\nu}}{4} (t - t_0) \exp\left[ -i \mathcal{E}_{\vec k_0} (t - t_0) \right] \sum_{s^{\prime\prime}} \\
&\hspace{1.95 cm} \bigg(F_a\, L_{2,1}^{+,s';+,s^{\prime\prime};\mu} L_{1,0}^{+,s^{\prime\prime};+,s;\nu}\label{eq:perturbative_propagator_dirac_equation_begin}\\
&\hspace{1.8 cm}+ F_b\, L_{2,1}^{+,s';+,s^{\prime\prime};\nu} L_{1,0}^{+,s^{\prime\prime};+,s;\mu} \\
&\hspace{1.8 cm}+ F_c\, L_{2,1}^{+,s';-,s^{\prime\prime};\nu} L_{1,0}^{-,s^{\prime\prime};+,s;\mu} \\
&\hspace{1.8 cm}+ F_d\, L_{2,1}^{+,s';-,s^{\prime\prime};\mu} L_{1,0}^{-,s^{\prime\prime};+,s;\nu} \bigg)\,,\label{eq:perturbative_propagator_dirac_equation_end}
\end{align}\label{eq:perturbative_propagator_dirac_equation}%
\end{subequations}%
with the prefactors
\begin{subequations}%
\begin{align}%
 F_a &= (\mathcal{E}_{\vec k_0} - \mathcal{E}_{\vec k_1} + k_l)^{-1}\\
 F_b &= (\mathcal{E}_{\vec k_0} - \mathcal{E}_{\vec k_1} - k_l)^{-1}\\
 F_c &= (\mathcal{E}_{\vec k_0} + \mathcal{E}_{\vec k_1} - k_l)^{-1}\\
 F_d &= (\mathcal{E}_{\vec k_0} + \mathcal{E}_{\vec k_1} + k_l)^{-1}\,.
\end{align}\label{eq:perturbative_prefactors_dirac_equation}%
\end{subequations}%
Note, that the lower integration limit of the $t_1$ integral in Eq. \eqref{eq:unintegrated_perturbative_propagator_dirac_equation} is only contributing non-resonant terms, which are neglected in Eq. \eqref{eq:perturbative_propagator_dirac_equation}.

\section{Second order Taylor expansion of the spin-dependent electron scattering matrix \label{sec:taylor_expansion_compton_tensor}}

For the terms in the last four lines in Eq. \eqref{eq:perturbative_propagator_dirac_equation} we define the expression
\begin{subequations}%
\begin{align}%
&M^{s',s;\mu\nu}=m \sqrt{\frac{\mathcal{E}_{\vec k_2}}{m}}
\sqrt{\frac{\mathcal{E}_{\vec k_0}}{m}} \sum_{s^{\prime\prime}} \\
&\hspace{1.95 cm} \bigg(F_a\, L_{2,1}^{+,s';+,s^{\prime\prime};\mu} L_{1,0}^{+,s^{\prime\prime};+,s;\nu}\\
&\hspace{1.8 cm}+ F_b\, L_{2,1}^{+,s';+,s^{\prime\prime};\nu} L_{1,0}^{+,s^{\prime\prime};+,s;\mu} \\
&\hspace{1.8 cm}+ F_c\, L_{2,1}^{+,s';-,s^{\prime\prime};\nu} L_{1,0}^{-,s^{\prime\prime};+,s;\mu} \\
&\hspace{1.8 cm}+ F_d\, L_{2,1}^{+,s';-,s^{\prime\prime};\mu} L_{1,0}^{-,s^{\prime\prime};+,s;\nu} \bigg)\,,
\end{align}\label{eq:spin_matrix_definition}%
\end{subequations}%
such that Eq. \eqref{eq:perturbative_propagator_dirac_equation} can be written as
\begin{multline}%
U_{2,0}^{+,s';+,s}(t,t_0) \approx -i \sqrt{\frac{m}{\mathcal{E}_{\vec k_2}}}
\sqrt{\frac{m}{\mathcal{E}_{\vec k_0}}} \frac{\textrm{e}^2 a_{\mu}^{\prime *} a_{\nu}}{4 m} M^{s',s;\mu\nu} \\
(t - t_0) \exp\left[ -i \mathcal{E}_{\vec k_0} (t - t_0) \right]. \label{eq:perturbative_propagator_taylor}
\end{multline}%
The matrix elements $M^{s',s,\mu\nu}$ in \eqref{eq:spin_matrix_definition} are functions of the photon momentum $k_l$ and the two transverse photon momenta $k_2$ and $k_3$. For the following calculation, we introduce the scaled parameters
\begin{equation}
 q_l = \frac{k_l}{m}\,,\qquad
 q_2 = \frac{k_2}{m}\,,\qquad
 q_3 = \frac{k_3}{m}\label{eq:reduced_taylor_momenta}
\end{equation}
and
\begin{equation}
\tilde q_3 = \frac{k_3 - m}{m} = q_3 - 1\,.\label{eq:shifted_k3_momentum}
\end{equation}
The Taylor expansion of $M^{s',s,\mu\nu}$ with respect to the three parameters \eqref{eq:reduced_taylor_momenta} is
\begin{widetext}
	\begin{subequations}%
		\begin{align}%
		M^{22} &= \left(1 + \frac{\sqrt{2}-1}{2}q_l^2 - q_2^2\right)\mathds{1} -\frac{i}{\sqrt{2}}\left[\left(\sqrt{2} - 1\right) + \frac{3 - 2\sqrt{2}}{2}\tilde q_3\right]q_l \sigma_y - \frac{i}{\sqrt{2}} q_l q_2 \sigma_z \label{eq:M_22} \\
		M^{23} &= - q_2 \mathds{1} + \left[ -\frac{i}{2} \sigma_x + \frac{i}{2} \tilde q_3 \sigma_x + \frac{i}{2} q_2 \sigma_y - \frac{i}{2} \sigma_z\right] q_l \label{eq:M_23}\\
		M^{32} &= - q_2 \mathds{1} + \left[ \phantom{-}\frac{i}{2} \sigma_x -\frac{i}{2} \tilde q_3 \sigma_x + \frac{i}{2} q_2 \sigma_y - \frac{i}{2} \sigma_z\right] q_l \label{eq:M_32}\\
		M^{33} &= \left[ -\tilde q_3 + \frac{1}{2} \tilde q_3^2 + \frac{\sqrt{2}-1}{2}q_l^2 + \frac{q_2^2}{2} \right]\mathds{1} - \frac{i}{\sqrt{2}}\left[ -1 + \frac{3 - 2 \sqrt{2}}{2}\tilde q_3 \right]q_l \sigma_y + i \frac{\sqrt{2} - 1}{\sqrt{2}}q_2 q_l \sigma_z \,. \label{eq:M_33}
		\end{align}\label{eq:taylor_expansion_compton_tensor}%
	\end{subequations}%
\end{widetext}
Here, we have accounted for all contributions up to the quadratic order in the expansion parameters $q_l$, $q_2$ and $q_3$ and their mixed orders. Note, that the Taylor expansion with respect to $q_l$ and $q_2$ is performed around their zero value $k_l=0$ and $k_2=0$, while the Taylor expansion with respect to $q_3$ is performed around the value $k_3=m$, to get an approximate expression in the vicinity around the initial and final momenta $\tilde {\vec p}_i$ and $\tilde {\vec p}_f$ (as defined in Eq. \eqref{eq:initial_electron_momentum} and \eqref{eq:final_electron_momentum}), about which the whole article is about. For this reason we have rewritten the electron momentum $q_3$ into the shifted momentum $\tilde q_3$ in Eq. \eqref{eq:shifted_k3_momentum}, where the Taylor expansion around the value $k_3=m$ corresponds to a Taylor expansion around $\tilde q_3=0$. We point out that the Taylor expanded matrix \eqref{eq:taylor_expansion_compton_tensor} shows the same matrix entries as the matrix (5) in reference \cite{ahrens_2017_spin_non_conservation}, with the addition that Eq. \eqref{eq:taylor_expansion_compton_tensor} also shows the second order terms of the Taylor expansion.

\section{Perturbative electron interaction with a quantized photon field\label{sec:qed_perturbation_theory}}

\subsection{Development of frame-fixed, quantized electron-photon Hamiltonian}

We now want to perform a similar perturbative procedure of the above appendix \ref{sec:dirac_perturbation_theory} for a system, where a single electron is interacting with a single photon and where particles are quantized in the context of a canonical quantization. Thus, we start by assuming the initial two particle excitation
\begin{equation}
 \Psi_i = c^{s\dagger}_{\vec p_i} a^{w\dagger}_{\vec k} \ket{0} \,,\label{eq:initial_state}
\end{equation}
where $c^{s\dagger}_{\vec p_i}$ is the electron creation operator with spin state $s$ and initial momentum $\vec p_i$ and $a^{w\dagger}_{\vec k}$ is the photon creation operator with polarization $w$ and momentum $\vec k$. The ket $\ket{0}$ is the quantum vacuum state with a zero number of electron and photon excitations. For the particle operators, we assume commutation relations $[\cdot,\cdot]$ for the photon particle operators and anti-commutation $\{\cdot,\cdot\}$ for the electron particle and anti-particle operators
\begin{subequations}%
\begin{align}%
 [a^{\lambda}_{\vec k},a^{\eta\dagger}_{\vec k'}]&=\{c^{\lambda}_{\vec k},c^{\eta\dagger}_{\vec k'}\}=\{d^{\lambda}_{\vec k},d^{\eta\dagger}_{\vec k'}\}=\delta_{\vec k,\vec k'} \delta_{\lambda,\eta}\\
 [a^{r}_{\vec k},a^{t}_{\vec k'}]&=\{c^{s}_{\vec k},c^{s'}_{\vec k'}\}=\{d^{s}_{\vec k},d^{s'}_{\vec k'}\}=0\,.
\end{align}%
We also assume that electron particle and anti-particle operators anti-commute with each other and photon operators commute with electron particle and anti-particle operators.%
\begin{alignat}{3}%
 \{c^{s}_{\vec k},d^{s'}_{\vec k'}\}&=\{c^{s}_{\vec k},d^{s'\dagger}_{\vec k'}\}&&=0\\
 [a^r_{\vec k},c^r_{\vec k'}]&=[a^r_{\vec k},c^{r\dagger}_{\vec k'}]&&=0\\
 [a^r_{\vec k},d^r_{\vec k'}]&=[a^r_{\vec k},d^{r\dagger}_{\vec k'}]&&=0
\end{alignat}\label{eq:particle_commutation_relations}%
\end{subequations}%
Our aim is to find the perturbative time evolution under the action
\begin{equation}
\mathcal{L} = \bar \Psi \left( i \gamma_\mu \partial^\mu - m \right) \Psi - \frac{1}{4} F^{\mu \nu} F_{\mu \nu} - \textrm{e} \bar \Psi \gamma_\mu \mathcal{A}^\mu \Psi \,,\label{eq:qed_lagrangian}
\end{equation}
with Dirac field $\Psi$, photon field $\mathcal{A}^\mu$ and electro-magnetic field tensor
\begin{equation}
 F^{\mu\nu} = \partial^\mu \mathcal{A}^\nu - \partial^\nu \mathcal{A}^\mu\,,
\end{equation}
with the derivative
\begin{equation}
 \partial^\mu = \frac{\partial}{\partial x_\mu}\,.
\end{equation}
The bar on top of $\Psi$ denotes the multiplication of its adjoint with $\gamma_0$, $\bar \Psi = \Psi^\dagger \gamma_0\,$. Regarding the QED Lagrangian \eqref{eq:qed_lagrangian}, we go along conventions from standard quantum field theory, see \cite{Peskin_Schroeder_1995_Quantum_Field_Theory,Landau_Lifshitz_1982_Quantum_Electrodynamics,Ryder_1986_Quantum_Field_Theory,Srednicki_2007_Quantum_Field_Theory,Halzen_Martin_1984_quarks_and_leptons,Weinberg_1995_quantum_theory_of_fields} for introduction.

The Lagrangian density in Eq. \eqref{eq:qed_lagrangian} implies the free Hamiltonian for electrons and their anti-particles \cite{schwabl_2000_advanced_quantum_mechanics}
\begin{equation}
 H_{\textrm{e}} = \sum_{\vec k, s} \mathcal{E}_{\vec k} \left( c_{\vec k}^{s\dagger} c^s_{\vec k} + d^{s\dagger}_{\vec k} d^s_{\vec k} \right) \label{eq:free_electron_Hamiltonian}
\end{equation}
as well as the free Hamiltonian for photons \cite{schwabl_2000_advanced_quantum_mechanics}
\begin{equation}
 H_{\textrm{p}} = \sum_{\vec k, r} k \, a^{r\dagger}_{\vec k} a^r_{\vec k} \,, \label{eq:free_photon_Hamiltonian}
\end{equation}
where $k = |\vec k|$ is the dispersion of light in vacuum. Of particular interest for the time evolution is the interaction part of the Hamiltonian from the Hamiltonian density
\begin{equation}
 \mathcal{H} = \Pi\, \dot \Psi + \Pi^\mu \dot {\mathcal{A}}_\mu - \mathcal{L}\,,
\end{equation}
where $\Pi$ and $\Pi^\mu$ are the conjugated momenta of the Dirac field and the photon field, respectively. The interaction part of the Lagrangian density \eqref{eq:qed_lagrangian} is
\begin{equation}
 \mathcal{L}_{\textrm{int}} = - \textrm{e} \bar \Psi \gamma_\mu \mathcal{A}^\mu \Psi \,,
\end{equation}
implying the interaction part of the Hamiltonian density
\begin{equation}
 \mathcal{H}_{\textrm{int}} = \textrm{e} \bar \Psi \gamma_\mu \mathcal{A}^\mu \Psi \,.\label{eq:hamilton_density}
\end{equation}
We denote the electron field operators $\Psi, \bar \Psi$ and photon field operator $\mathcal{A}_\mu$ by
\begin{subequations}%
\begin{align}%
 \Psi(\vec x,t) &= \sum_{\vec k, s} \left( c^s_{\vec k} u^s_{\vec k} e^{- i k  \cdot x} + d^{s\dagger}_{\vec k} v^s_{\vec k} e^{i k \cdot x} \right) \label{eq:field_operator_electrons}\\
 \bar \Psi(\vec x,t) &= \sum_{\vec k, s} \left( c^{s\dagger}_{\vec k} \bar u^s_{\vec k} e^{i k \cdot x} + d^s_{\vec k} \bar v^s_{\vec k} e^{- i k \cdot x} \right) \\
 \mathcal{A}_\mu(\vec x,t) &= \sum_{\vec k, r} \left( \epsilon^{(r)}_{\mu,\vec k} a^{r}_{\vec k} e^{- i k \cdot x} + \epsilon^{(r) *}_{\mu,\vec k} a^{r\dagger}_{\vec k} e^{i k \cdot x} \right)\,,\label{eq:field_operator_photons}
\end{align}\label{eq:field_operators}%
\end{subequations}%
where $u^s_{\vec k}$ and $v^s_{\vec k}$ are the bi-spinors \eqref{eq:bi-spinors} and $\epsilon^{(r)}_{\mu,\vec k}$ are the four ($r$ index) four-polarization ($\mu$ index) vectors of the photon field.
Inserting the definitions \eqref{eq:field_operators} in the interaction part of the Hamilton density \eqref{eq:hamilton_density} yields the interaction Hamiltonian
\begin{subequations}%
\begin{align}%
 H_{\textrm{int}} &= \textrm{e} \int d^3 x \bar \Psi \gamma_\mu \mathcal{A}^\mu \Psi = \textrm{e} \sum_{\vec k, \vec k' \atop s, s', r} \Bigg[\\
 & \phantom{+}\,\ \left( \bar u^s_{\vec k} \slashed \epsilon^{(r)}_{\vec k - \vec k'} u^{s'}_{\vec k'}\right) c^{s\dagger}_{\vec k} c^{s'}_{\vec k'} a^r_{\vec k - \vec k'}\\
 &          +  \left( \bar u^s_{\vec k} \slashed \epsilon^{(r)*}_{- \vec k + \vec k'} u^{s'}_{\vec k'}\right) c^{s\dagger}_{\vec k} c^{s'}_{\vec k'} a^{r\dagger}_{- \vec k + \vec k'}\\
 &          +  \left( \bar v^s_{-\vec k} \slashed \epsilon^{(r)}_{\vec k - \vec k'} u^{s'}_{\vec k'}\right) d^s_{-\vec k} c^{s'}_{\vec k'} a^r_{\vec k - \vec k'}\\
 &          +  \left( \bar v^s_{-\vec k} \slashed \epsilon^{(r)*}_{- \vec k + \vec k'} u^{s'}_{\vec k'}\right) d^s_{-\vec k} c^{s'}_{\vec k'} a^{r\dagger}_{- \vec k + \vec k'}\\
 &          +  \left( \bar u^s_{\vec k} \slashed \epsilon^{(r)}_{\vec k - \vec k'} v^{s'}_{-\vec k'}\right) c^{s\dagger}_{\vec k} d^{s'\dagger}_{- \vec k'} a^r_{\vec k - \vec k'}\\
 &          +  \left( \bar u^s_{\vec k} \slashed \epsilon^{(r)*}_{- \vec k + \vec k'} v^{s'}_{-\vec k'}\right) c^{s\dagger}_{\vec k} d^{s'\dagger}_{- \vec k'} a^{r\dagger}_{- \vec k + \vec k'}\\
 &          +  \left( \bar v^s_{-\vec k} \slashed \epsilon^{(r)}_{\vec k - \vec k'} v^{s'}_{-\vec k'}\right) d^s_{-\vec k} d^{s'\dagger}_{- \vec k'} a^r_{\vec k - \vec k'}\\
 &          +  \left( \bar v^s_{-\vec k} \slashed \epsilon^{(r)*}_{- \vec k + \vec k'} v^{s'}_{-\vec k'}\right) d^s_{-\vec k} d^{s'\dagger}_{- \vec k'} a^{r\dagger}_{- \vec k + \vec k'}\Bigg]\,,
\end{align}\label{eq:interaction_hamiltonian}%
\end{subequations}%
where we are using the Feynman slash notion $\slashed{\epsilon} = \epsilon_\mu \gamma^\mu$ for abbreviation of the contraction of the Dirac gamma matrices with a four-vector. The obtained interaction Hamiltonian \eqref{eq:interaction_hamiltonian}, together with the free Hamiltonians \eqref{eq:free_electron_Hamiltonian} and \eqref{eq:free_photon_Hamiltonian} determine the time evolution of vacuum excitations in the Schr\"odinger picture by
\begin{equation}
 i \dot \Psi = H \Psi\,,\label{eq:time_evolution}
\end{equation}
with
\begin{equation}
 H = H_\textrm{e} + H_\textrm{p} + H_{\textrm{int}} \,.
\end{equation}

\subsection{Perturbative derivation with quantized Hamiltonian}

From Eq. \eqref{eq:time_evolution} one can write the time evolution in form of a Dyson series, whose second order interaction term reads
\begin{multline}
 \mathcal{U}(t,t_0) = \frac{1}{i^2} \int_{t_0}^t d t_2 \int_{t_0}^{t_2} d t_1 \\
 \times \mathcal{U}_0(t,t_2) H_{\textrm{int}} \mathcal{U}_0(t_2,t_1) H_{\textrm{int}} \mathcal{U}_0(t_1,t_0)\,,\label{eq:second_order_perturbation_theory}
\end{multline}
which is formally similar to the second order perturbation of the single particle description \eqref{eq:dirac_second_order_perturbation}.
Here, $\mathcal{U}_0(t,t_0)$ is the free propagation
\begin{equation}
  \mathcal{U}_0(t, t_0) = \exp\left[-i ( H_{\textrm{e}} + H_{\textrm{p}})(t-t_0)\right]\label{eq:free_propagator}
\end{equation}
of electrons, their anti-particles and photons. The first order contributions of the Dyson series are not considered, since they do not contain resonant terms due to energy and momentum conservation. For the initial state $\Psi_i$ in Eq. \eqref{eq:initial_state} we obtain from Eq. \eqref{eq:free_propagator}
\begin{equation}
  \mathcal{U}_0(t_1, t_0) = \exp\left[-i (\mathcal{E}_{\vec p_i} + k)(t_1-t_0)\right]\label{eq:free_propagation_psi_i}
\end{equation}
for the time interval $[t_0,t_1]$ of the first free quantum state propagation in the second order perturbation \eqref{eq:second_order_perturbation_theory}. Note, that while Eq. \eqref{eq:free_propagator} is an operator equation, the expressions in Eq. \eqref{eq:free_propagation_psi_i} and later in the text also the Eqs. \eqref{eq:free_propagation_intermediate} and \eqref{eq:free_propagation_final} are expressions where the operators have been acting at the operators of the quantum states and turned into ordinary complex numbers by the eigenvalue operations of the operators.

The first action of $H_{\textrm{int}}$ on $\Psi_i$ results in the intermediate states
\begin{subequations}%
\begin{align}%
 \Psi_a &= c^{s''\dagger}_{\vec p_i + \vec k} \ket{0} \\
 \Psi_b &= \ c^{s''\dagger}_{\vec p_i - \vec k'} a^{w\dagger}_{\vec k} a^{r\dagger}_{\vec k'} \ket{0} \\
 \Psi_c &= c^{s\dagger}_{\vec p_i} c^{s'\dagger}_{\vec p_f} d^{s''\dagger}_{-\vec p_i + \vec k'} \ket{0} \\
 \Psi_d &= \ c^{s\dagger}_{\vec p_i} c^{s'\dagger}_{\vec p_f} d^{s''\dagger}_{-\vec p_i - \vec k} a^{w\dagger}_{\vec k} a^{r\dagger}_{\vec k'} \ket{0} \,,
\end{align}\label{eq:intermediate_states}%
\end{subequations}%
where the interaction term
\begin{subequations}%
\begin{align}%
&\left( \bar u^{s''}_{\vec p_i + \vec k} \slashed \epsilon^{(w)}_{\vec k} u^{s}_{\vec p_i} \right)
c^{s''\dagger}_{\vec p_i + \vec k} c^{s}_{\vec p_i} a^w_{\vec k}\textrm{ maps to }\Psi_a\\
&\left( \bar u^{s''}_{\vec p_i - \vec k'} \slashed \epsilon^{(r)*}_{\vec k'} u^{s}_{\vec p_i} \right)
c^{s''\dagger}_{\vec p_i - \vec k'} c^{s}_{\vec p_i} a^{r\dagger}_{\vec k'} \textrm{ maps to }\Psi_b\\
&\left( \bar u^{s'}_{\vec p_f} \slashed \epsilon^{(w)}_{\vec k} v^{s''}_{-\vec p_i + \vec k'} \right) 
c^{s'\dagger}_{\vec p_f} d^{s''\dagger}_{-\vec p_i + \vec k'} a^w_{\vec k}\textrm{ maps to }\Psi_c\\
&\left( \bar u^{s'}_{\vec p_f} \slashed \epsilon^{(r)*}_{\vec k'} v^{s''}_{-\vec p_i - \vec k} \right)
c^{s'\dagger}_{\vec p_f} d^{s''\dagger}_{-\vec p_i - \vec k} a^{r\dagger}_{\vec k'}\textrm{ maps to }\Psi_d,
\end{align}\label{eq:first_interaction_terms}%
\end{subequations}%
as illustrated in Fig. \ref{fig:electron_photon_states}. In correspondence, for the time interval $[t_1,t_2]$ of the second free propagation in \eqref{eq:second_order_perturbation_theory} one obtains the free propagation
\begin{align}
\mathcal{U}_0(t_2, t_1) &= \exp[-i ( \mathcal{E}_{\vec p_i + \vec k})(t_2-t_1)]\label{eq:free_propagation_intermediate}\\
\mathcal{U}_0(t_2, t_1) &= \exp[-i ( \mathcal{E}_{\vec p_i - \vec k'} + k + k')(t_2-t_1)]\nonumber\\
\mathcal{U}_0(t_2, t_1) &= \exp[-i ( \mathcal{E}_{\vec p_i} + \mathcal{E}_{\vec p_f} + \mathcal{E}_{-\vec p_i + \vec k'})(t_2-t_1)] \nonumber\\
\mathcal{U}_0(t_2, t_1) &= \exp[-i ( \mathcal{E}_{\vec p_i} + \mathcal{E}_{\vec p_f} + \mathcal{E}_{-\vec p_i - \vec k} + k + k')\nonumber\\
&\hspace{4.6 cm}\cdot(t_2-t_1)]\nonumber
\end{align}
for $\Psi_a$, $\Psi_b$, $\Psi_c$ and $\Psi_d$, respectively. For the following second interaction $H_{\textrm{int}}$ in Eq. \eqref{eq:second_order_perturbation_theory} only terms are relevant which fulfill energy conservation, as all other contributions will oscillate in off-resonant Rabi cycles of low amplitude. Particle excitations different than
\begin{equation}
 \Psi_f = c^{s'\dagger}_{\vec p_f} a^{r\dagger}_{\vec k'} \ket{0} \,,\label{eq:final_state}
\end{equation}
are therefore not possible for asymptotically long times, with the final electron momentum $\vec p_f = \vec p_i + \vec k - \vec k'$ and photon momentum $\vec k'$. The corresponding energy conservation relation of the constituting particles displays as
\begin{equation}
 \mathcal{E}_{\vec p_i} + k = \mathcal{E}_{\vec p_f} + k'\,.\label{eq:energy_conservation}
\end{equation}
According to the above considerations, the only contributions in the interaction Hamiltonian \eqref{eq:interaction_hamiltonian} that map back to the final state $\Psi_f$ are
\begin{subequations}%
\begin{align}%
&\left( \bar u^{s'}_{\vec p_f} \slashed \epsilon^{(r)*}_{\vec k'} u^{s''}_{\vec p_i + \vec k} \right)
c^{s'\dagger}_{\vec p_f} c^{s''}_{\vec p_i + \vec k} a^{r\dagger}_{\vec k'}
\textrm{ from }\Psi_a\\
&\left( \bar u^{s'}_{\vec p_f} \slashed \epsilon^{(w)}_{\vec k} u^{s''}_{\vec p_i - \vec k'} \right)
c^{s'\dagger}_{\vec p_f} c^{s''}_{\vec p_i - \vec k'} a^{w}_{\vec k} \textrm{ from }\Psi_b\\
&\left( \bar v^{s''}_{-\vec p_i + \vec k'} \slashed \epsilon^{(r)*}_{\vec k'} u^{s}_{\vec p_i} \right)
d^{s''}_{-\vec p_i + \vec k'} c^{s}_{\vec p_i} a^{r\dagger}_{\vec k'} \textrm{ from }\Psi_c\\
&\left( \bar v^{s''}_{-\vec p_i - \vec k} \slashed \epsilon^{(w)}_{\vec k} u^{s}_{\vec p_i} \right)
d^{s''}_{-\vec p_i - \vec k} c^{s}_{\vec p_i} a^{w}_{\vec k} \textrm{ from }\Psi_d\,.
\end{align}\label{eq:second_interaction_terms}%
\end{subequations}%
The free propagation \eqref{eq:free_propagator} of the final state $\Psi_f$ evaluates to
\begin{subequations}%
\begin{align}%
  \mathcal{U}_0(t, t_2) &= \exp\left[-i (\mathcal{E}_{\vec p_f} + k')(t-t_2)\right]\\
              &= \exp\left[-i (\mathcal{E}_{\vec p_i} + k)(t-t_2)\right]
\end{align}\label{eq:free_propagation_final}%
\end{subequations}%
and it's phase oscillates with the same frequency as the free propagation of $\Psi_i$ in Eq. \eqref{eq:free_propagation_psi_i}, due to the energy conservation relation \eqref{eq:energy_conservation}. Consequently, the oscillations of the perturbative contribution of the propagator \eqref{eq:second_order_perturbation_theory} with respect to the integration variables $t_1$ and $t_2$ oscillate in the exponential with the factor
\begin{subequations}%
\begin{align}%
 &-i(\mathcal{E}_{\vec p_i + \vec k} - \mathcal{E}_{\vec p_i} - k)(t_2 - t_1)\\
 &-i(\mathcal{E}_{\vec p_i - \vec k'} - \mathcal{E}_{\vec p_i} + k')(t_2 - t_1)\\
 &-i(\mathcal{E}_{-\vec p_i + \vec k'} + \mathcal{E}_{\vec p_i} - k')(t_2 - t_1)\\
 &-i(\mathcal{E}_{-\vec p_i - \vec k} + \mathcal{E}_{\vec p_i} + k)(t_2 - t_1)
\end{align}\label{eq:intermediate_time_phase_terms}%
\end{subequations}%
for $\Psi_a$, $\Psi_b$, $\Psi_c$ and $\Psi_d$, respectively. Note that Eq. \eqref{eq:energy_conservation} has been substituted, to arrive at \eqref{eq:intermediate_time_phase_terms}.

For the upper integration limit of the integral with respect to $t_1$ in Eq. \eqref{eq:second_order_perturbation_theory}, the phase terms \eqref{eq:intermediate_time_phase_terms} are canceling to zero, such that the integration with respect to $t_2$ will be independent of $t_2$, resulting in a solution which is growing linear in time, similar to the perturbative single particle calculation in appendix \ref{sec:dirac_perturbation_theory}.  In accordance, the integration with respect to $t_1$ yields the prefactors
\begin{subequations}%
\begin{align}%
 \mathcal{F}_a &= (\mathcal{E}_{\vec p_i} - \mathcal{E}_{\vec p_i + \vec k} + k)^{-1}\\
 \mathcal{F}_b &= (\mathcal{E}_{\vec p_i} - \mathcal{E}_{\vec p_i - \vec k'} - k')^{-1}\\
 \mathcal{F}_c &= (\mathcal{E}_{\vec p_i} + \mathcal{E}_{\vec p_i - \vec k'} - k')^{-1}\label{eq:perturbative_prefactors_c}\\
 \mathcal{F}_d &= (\mathcal{E}_{\vec p_i} + \mathcal{E}_{\vec p_i + \vec k} + k)^{-1}\,.\label{eq:perturbative_prefactors_d}
\end{align}\label{eq:perturbative_prefactors}%
\end{subequations}%
in Eq. \eqref{eq:second_order_perturbation_theory}, which are the analogon to the prefactors \eqref{eq:perturbative_prefactors_dirac_equation}. Note, that we accounted for an additional minus sign for Eqs. \eqref{eq:perturbative_prefactors_c} and \eqref{eq:perturbative_prefactors_d} due to the commutation relations \eqref{eq:particle_commutation_relations} of the additional virtual electron-positron pair and we multiplied all terms with another factor $i^{-1}$ for ease of notion. We also made use of $\mathcal{E}_{\vec p} = \mathcal{E}_{-\vec p}$ for the determination of the factors \eqref{eq:perturbative_prefactors} from \eqref{eq:intermediate_time_phase_terms}.

Taking the prefactors \eqref{eq:perturbative_prefactors}, together with the corresponding interaction matrix elements in \eqref{eq:first_interaction_terms} and \eqref{eq:second_interaction_terms}
and substituting them into the propagator \eqref{eq:second_order_perturbation_theory} results in the expression
\begin{align}
\mathcal{U}^{s',s;r,w}(t,t_0) &=-i \textrm{e}^2 (t - t_0) \exp\left[-i (\mathcal{E}_{\vec p_i} + k)(t-t_0)\right]\nonumber\\
\times \sum_{s''} \Bigg[& \mathcal{F}_a \left( \bar u^{s'}_{\vec p_f} \slashed \epsilon^{(r)*}_{\vec k'} u^{s''}_{\vec p_i + \vec k} \right)\left( \bar u^{s''}_{\vec p_i + \vec k} \slashed \epsilon^{(w)}_{\vec k} u^{s}_{\vec p_i} \right)\nonumber\\
+ & \mathcal{F}_b \left( \bar u^{s'}_{\vec p_f} \slashed \epsilon^{(w)}_{\vec k} u^{s''}_{\vec p_i - \vec k'} \right)\left( \bar u^{s''}_{\vec p_i - \vec k'} \slashed \epsilon^{(r)*}_{\vec k'} u^{s}_{\vec p_i} \right)\nonumber\\
+ & \mathcal{F}_c \left( \bar u^{s'}_{\vec p_f} \slashed \epsilon^{(w)}_{\vec k} v^{s''}_{-\vec p_i + \vec k'} \right)\left( \bar v^{s''}_{-\vec p_i + \vec k'} \slashed \epsilon^{(r)*}_{\vec k'} u^{s}_{\vec p_i} \right)\nonumber\\
+ & \mathcal{F}_d \left( \bar u^{s'}_{\vec p_f} \slashed \epsilon^{(r)*}_{\vec k'} v^{s''}_{-\vec p_i - \vec k} \right)\left( \bar v^{s''}_{-\vec p_i - \vec k} \slashed \epsilon^{(w)}_{\vec k} u^{s}_{\vec p_i} \right)\Bigg]\,.\label{eq:full_and_exact_propagator}
\end{align}
In technical terms, the propagator \eqref{eq:full_and_exact_propagator} should be understood in the following way: The operator $\mathcal{U}(t,t_0)$ in Eq. \eqref{eq:second_order_perturbation_theory} is applied at the initial quantum state $c^{s\dagger}_{\vec p_i} a^{w\dagger}_{\vec k} \ket{0}$ in Eq. \eqref{eq:initial_state} and the propagation matrix \eqref{eq:full_and_exact_propagator} is determining the amplitude of the final states $c^{s'\dagger}_{\vec p_f} a^{r\dagger}_{\vec k'} \ket{0}$ of Eq. \eqref{eq:final_state}. These final states are the only relevant, resonant states, which are seen as non-vanishing contributions after long times $t$. In this context we conclude the approximate relation
\begin{equation}
 \mathcal{U}(t,t_0) \, c^{s\dagger}_{\vec p_i} a^{w\dagger}_{\vec k} \ket{0} \approx  \sum_{s',r} \mathcal{U}^{s',s;r,w}(t,t_0) \,c^{s'\dagger}_{\vec p_f} a^{r\dagger}_{\vec k'} \ket{0}\label{eq:propagator_mapping}
\end{equation}
for a perturbative electron-photon interaction. The spin and polarization properties of $\Psi_f$ are determined by matrix entries as in Eq. \eqref{eq:taylor_expansion_compton_tensor} for the scattering scenario as described in section \ref{sec:conceptual_remarks} and \ref{sec:theory_description}. Correspondingly, we find
\begin{subequations}%
\begin{align}%
 \mathcal{U}(t,t_0) \, c^{\searrow\dagger}_{\tilde {\vec p}_i} a^{3\dagger}_{\vec k_l} \ket{0} &\approx  f(t,t_0) \,c^{\nwarrow\dagger}_{\tilde {\vec p}_f} a^{\textrm{L}\dagger}_{-\vec k_l} \ket{0} \label{eq:compton_process_A}\\
 \mathcal{U}(t,t_0) \, c^{\nwarrow\dagger}_{\tilde {\vec p}_i} a^{3\dagger}_{\vec k_l} \ket{0} &\approx  f(t,t_0) \,c^{\searrow\dagger}_{\tilde {\vec p}_f} a^{\textrm{R}\dagger}_{-\vec k_l} \ket{0}\,,
\end{align}\label{eq:compton_process}%
\end{subequations}%
with the time dependent prefactor
\begin{equation}
f(t,t_0)=\textrm{e}^2 (t-t_0)\exp[-i(\mathcal{E}_{\vec p_i}+k)(t - t_0)]\,.
\end{equation}
Here, the tilted spin electron creation and annihilation operators can be expressed in terms of the spin up and spin down electron creation operators $c^{\uparrow\dagger}_{\vec p}$ and $c^{\downarrow\dagger}_{\vec p}$ and the spin states $s^\searrow$ and $s^\nwarrow$ of Eq. \eqref{eq:tilted_spin_states} and Eq. \eqref{eq:tilted_spin_states_alternative_definition} by
\begin{subequations}%
\begin{align}%
 c^{\searrow\dagger}_{\vec p} &= s_1^\searrow c^{\uparrow\dagger}_{\vec p} + s_2^\searrow c^{\downarrow\dagger}_{\vec p}\\
 c^{\nwarrow\dagger}_{\vec p} &= s_1^\nwarrow c^{\uparrow\dagger}_{\vec p} + s_2^\nwarrow c^{\downarrow\dagger}_{\vec p}\,.
\end{align}%
\end{subequations}%
Also, the left and right handed photon creation operators in \eqref{eq:compton_process} are defined by
\begin{subequations}%
\begin{align}%
 a^{L\dagger}_{\vec k_l} &= a^{2\dagger}_{\vec k_l} - i a^{3\dagger}_{\vec k_l}\\
 a^{R\dagger}_{\vec k_l} &= a^{2\dagger}_{\vec k_l} + i a^{3\dagger}_{\vec k_l}\\
 a^{L\dagger}_{-\vec k_l} &= a^{2\dagger}_{-\vec k_l} + i a^{3\dagger}_{-\vec k_l}\\
 a^{R\dagger}_{-\vec k_l} &= a^{2\dagger}_{-\vec k_l} - i a^{3\dagger}_{-\vec k_l}\,.
\end{align}\label{eq:circular_polarization}%
\end{subequations}%
We point out that Eq. \eqref{eq:compton_process} is the corresponding expression to Eq. (19) in Ref. \cite{ahrens_2017_spin_non_conservation}. However, in contrast to Ref. \cite{ahrens_2017_spin_non_conservation}, the definitions \eqref{eq:circular_polarization} contain consistent helicities of the photons which are propagating in the $x$ or $-x$ direction, in contrast to the left and right circular polarization introduced in reference \cite{ahrens_2017_spin_non_conservation}. Also note that reference \cite{ahrens_2017_spin_non_conservation} is not accounting for the complex conjugation of the outgoing photon polarization in the Compton scattering formula \eqref{eq:compton_tensor} (see for example \cite{Peskin_Schroeder_1995_Quantum_Field_Theory}). In this work both issues are accounted for.

\subsection{Identification with Compton scattering formula from quantum field theory\label{sec:compton_tensor_identification}}

The photon polarization dependent electron spin coupling matrix in Eq. \eqref{eq:full_and_exact_propagator} consists of the components
\begin{subequations}%
\begin{align}%
&\sum_{s''} \mathcal{F}_a \left( \bar u^{s'}_{\vec p_f} \slashed \epsilon^{(r)*}_{\vec k'} u^{s''}_{\vec p_i + \vec k} \right)\left( \bar u^{s''}_{\vec p_i + \vec k} \slashed \epsilon^{(w)}_{\vec k} u^{s}_{\vec p_i} \right)\label{eq:perturbative_transition_amplitude_a}\\
&\sum_{s''} \mathcal{F}_b \left( \bar u^{s'}_{\vec p_f} \slashed \epsilon^{(w)}_{\vec k} u^{s''}_{\vec p_i - \vec k'} \right)\left( \bar u^{s''}_{\vec p_i - \vec k'} \slashed \epsilon^{(r)*}_{\vec k'} u^{s}_{\vec p_i} \right)\label{eq:perturbative_transition_amplitude_b}\\
&\sum_{s''} \mathcal{F}_c \left( \bar u^{s'}_{\vec p_f} \slashed \epsilon^{(w)}_{\vec k} v^{s''}_{-\vec p_i + \vec k'} \right)\left( \bar v^{s''}_{-\vec p_i + \vec k'} \slashed \epsilon^{(r)*}_{\vec k'} u^{s}_{\vec p_i} \right)\label{eq:perturbative_transition_amplitude_c}\\
&\sum_{s''} \mathcal{F}_d \left( \bar u^{s'}_{\vec p_f} \slashed \epsilon^{(r)*}_{\vec k'} v^{s''}_{-\vec p_i - \vec k} \right)\left( \bar v^{s''}_{-\vec p_i - \vec k} \slashed \epsilon^{(w)}_{\vec k} u^{s}_{\vec p_i} \right)\label{eq:perturbative_transition_amplitude_d}
\end{align}\label{eq:perturbative_transition_amplitudes}%
\end{subequations}%
where each line corresponds to the intermediate quantum states \eqref{eq:intermediate_states} and the corresponding spin and polarization dependent matrix elements as well as prefactors have been denoted in equations \eqref{eq:first_interaction_terms}, \eqref{eq:second_interaction_terms} and \eqref{eq:perturbative_prefactors}, respectively. These expressions can be further simplified to appear as final $S$-matrix expressions in quantum field theory. First we can substitute the identities
\begin{subequations}%
\begin{align}%
 \sum _{s} u^{s}_{\vec p} \bar u^{s}_{\vec p} &= \frac{\slashed p + m}{2 \mathcal{E}_{\vec p}}\\
 \sum _{s} v^{s}_{\vec p} \bar v^{s}_{\vec p} &= \frac{\slashed p - m}{2 \mathcal{E}_{\vec p}}\,,
\end{align}%
\end{subequations}%
into the expressions \eqref{eq:perturbative_transition_amplitudes}, resulting in
\begin{subequations}%
\begin{align}%
& (\mathcal{F}_a - \mathcal{F}_d) \bar u^{s'}_{\vec p_f} \slashed \epsilon^{(r)*}_{\vec k'} \frac{ \slashed p_i + \slashed k + m }{2 \mathcal{E}_{\vec p_i + \vec k}} \slashed \epsilon^{(w)}_{\vec k} u^{s}_{\vec p_i}\label{eq:photon_absorption_term}\\
& (\mathcal{F}_b - \mathcal{F}_c) \bar u^{s'}_{\vec p_f} \slashed \epsilon^{(w)}_{\vec k} \frac{ \slashed p_i - \slashed k' + m }{2 \mathcal{E}_{\vec p_i - \vec k'}} \slashed \epsilon^{(r)*}_{\vec k'} u^{s}_{\vec p_i}\,.\label{eq:photon_emission_term}
\end{align}%
\end{subequations}%
Eqs. \eqref{eq:perturbative_transition_amplitude_a} and \eqref{eq:perturbative_transition_amplitude_d}, as well as Eqs. \eqref{eq:perturbative_transition_amplitude_b} and \eqref{eq:perturbative_transition_amplitude_c} are summed up into Eq. \eqref{eq:photon_absorption_term} and Eq. \eqref{eq:photon_emission_term}, respectively. In Eq. \eqref{eq:photon_absorption_term} we can simplify
\begin{equation}
 \frac{\mathcal{F}_a - \mathcal{F}_d}{2 \mathcal{E}_{\vec p_i + \vec k}} = \frac{1}{(\mathcal{E}_{\vec p_i} + k)^2 - (\mathcal{E}_{\vec p_i + \vec k})^2} = \frac{1}{2 p_i \cdot k}
\end{equation}
and similarly in Eq. \eqref{eq:photon_emission_term} we can simplify
\begin{equation}
 \frac{\mathcal{F}_b - \mathcal{F}_c}{2 \mathcal{E}_{\vec p_i - \vec k'}} = \frac{1}{(\mathcal{E}_{\vec p_i} - k')^2 - (\mathcal{E}_{\vec p_i - \vec k'})^2} = - \frac{1}{2 p_i \cdot k'}\,.
\end{equation}
Summing up also Eqs. \eqref{eq:photon_absorption_term} and \eqref{eq:photon_emission_term} finally results in the Compton scattering formula
\begin{equation}
 \bar u^{s'}_{\vec p_f} \left( \slashed \epsilon^{(r)*}_{\vec k'} \frac{ \slashed p_i + \slashed k + m }{2 p_i \cdot k} \slashed \epsilon^{(w)}_{\vec k} - \slashed \epsilon^{(w)}_{\vec k} \frac{ \slashed p_i - \slashed k' + m }{2 p_i \cdot k'} \slashed \epsilon^{(r)*}_{\vec k'}\right) u^{s}_{\vec p_i}\,.\label{eq:compton_tensor}
\end{equation}
Note, that Eq. \eqref{eq:compton_tensor} is a rewritten version of Eq. \eqref{eq:perturbative_transition_amplitudes}, which in turn can be associated with the perturbative solution in Eqs.  \eqref{eq:perturbative_propagator_dirac_equation_begin} till \eqref{eq:perturbative_propagator_dirac_equation_end} for spin-dependent electron diffraction. A difference between both expressions is, that \eqref{eq:perturbative_propagator_dirac_equation} is constructed from a standing light wave situation with $\vec k=-\vec k'$, whereas in the expressions in \eqref{eq:perturbative_transition_amplitudes} the wave vectors $\vec k$ and $\vec k'$ could be chosen independently. Also, we have used the abbreviation $L_{n,n'}^{\gamma,s;\gamma',s';\mu}$ for abbreviating the matrix elements \eqref{eq:spinor_matrix_contractions} in \eqref{eq:perturbative_propagator_dirac_equation}. We point out that our calculations show explicitly that he spin and polarization dependent interaction of an electron with a photon in the process of Compton scattering  \eqref{eq:compton_tensor} is matching to the perturbative description of spin-dependent diffraction dynamics of an electron in an external potential of two plane waves \eqref{eq:perturbative_propagator_dirac_equation}. In other words, the spin/polarization properties in Compton scattering and in electron diffraction of the Kapitza-Dirac effect are of identical form, only the interpretation of the associated process is different, depending on the scenario of consideration (ie. whether the scenario is Compton scattering or electron diffraction).

Further details about the calculation in this section can be found in the literature under the name 'old-fashioned perturbation theory', see for example \cite{Halzen_Martin_1984_quarks_and_leptons,Weinberg_1995_quantum_theory_of_fields}.

\section{Spin-projection of propagator\label{sec:tilted_spin_propagation}}

Assume, that $U^{+,+}$ is a complex $2\times 2$ matrix, which is representing the propagation of a two-component spinor. Then the following matrix elements can be written as
\begin{widetext}%
\begin{subequations}%
\begin{align}%
 \Braket{s^\searrow | U_{a,b}^{+;+} | s^\searrow } &= \frac{1}{\sqrt{8}} \bigg[ ( 1 - \sqrt{2}) U_{a,b}^{+,\uparrow;+,\uparrow} - U_{a,b}^{+,\uparrow;+,\downarrow} - U_{a,b}^{+,\downarrow;+,\uparrow} - ( 1 + \sqrt{2}) U_{a,b}^{+,\downarrow;+,\downarrow} \bigg] \\
 \Braket{s^\searrow | U_{a,b}^{+;+} | s^\nwarrow } &= \frac{1}{\sqrt{8}} \bigg[ - U_{a,b}^{+,\uparrow;+,\uparrow} - ( 1 - \sqrt{2}) U_{a,b}^{+,\uparrow;+,\downarrow} - ( 1 + \sqrt{2}) U_{a,b}^{+,\downarrow;+,\uparrow} + U_{a,b}^{+,\downarrow;+,\downarrow} \bigg] \\
 \Braket{s^\nwarrow | U_{a,b}^{+;+} | s^\searrow } &= \frac{1}{\sqrt{8}} \bigg[ - U_{a,b}^{+,\uparrow;+,\uparrow} - ( 1 + \sqrt{2}) U_{a,b}^{+,\uparrow;+,\downarrow} - ( 1 - \sqrt{2}) U_{a,b}^{+,\downarrow;+,\uparrow} + U_{a,b}^{+,\downarrow;+,\downarrow} \bigg] \\
 \Braket{s^\nwarrow | U_{a,b}^{+;+} | s^\nwarrow } &= \frac{1}{\sqrt{8}} \bigg[ -( 1 + \sqrt{2}) U_{a,b}^{+,\uparrow;+,\uparrow} + U_{a,b}^{+,\uparrow;+,\downarrow} + U_{a,b}^{+,\downarrow;+,\uparrow} + ( 1 - \sqrt{2}) U_{a,b}^{+,\downarrow;+,\downarrow} \bigg]\,.
\end{align}\label{eq:tilted_propagator_spin_projections}%
\end{subequations}%
\end{widetext}%

\bibliography{bibliography}

\end{document}